\documentclass[floats,floatfix,showpacs,amssymb,prd,twocolumn,superscriptaddress,nofootinbib]{revtex4-1}

\newlength{\figw} 
\usepackage{graphicx,epsf, epsfig, amssymb}
\usepackage{bm}
\usepackage{longtable,tabularx}
\usepackage{color}
\usepackage[breaklinks]{hyperref}
\usepackage{amsfonts,amsmath,amssymb,mathrsfs,gensymb}
\usepackage{natbib}
\usepackage{aas_macros}
\usepackage{array}
\usepackage{multirow}
\usepackage{rotating,array}

\newcommand\optional[1]{}
\def\be{\begin{equation}}
\def\ee{\end{equation}}
\def\beq{\begin{eqnarray}}
\def\eeq{\end{eqnarray}}
\usepackage{comment}

\newcommand{\sun}{\ensuremath{\odot}}

\begin{document}

\title{Resonant-plane locking and spin alignment in stellar-mass
  black-hole binaries:\\a diagnostic of compact-binary formation}

\author{Davide Gerosa}
\email{dgerosa@olemiss.edu}

\affiliation{Department of Physics and Astronomy, The University of
Mississippi, University, MS 38677, USA}
\affiliation{Dipartimento di Fisica, Universit\'a Degli Studi di Milano, Via Celoria, 16, Milano, 20133, Italy}

\author{Michael Kesden}
\email{mhk10@nyu.edu}
\affiliation{Center for Cosmology and Particle Physics, New York University, 4 Washington Pl., New York, NY 10003, USA}

\author{Emanuele Berti}
\email{berti@phy.olemiss.edu}
\affiliation{Department of Physics and Astronomy, The University of
Mississippi, University, MS 38677, USA}
\affiliation{California Institute of Technology, Pasadena, CA 91109, USA}

\author{Richard O'Shaughnessy}
\email{oshaughn@gravity.phys.uwm.edu}
\affiliation{Center for Gravitation and Cosmology, University of Wisconsin-Milwaukee, Milwaukee, WI 53211, USA}

\author{Ulrich Sperhake}
\email{sperhake@tapir.caltech.edu}
\affiliation{Department of Applied Mathematics and Theoretical Physics, Centre for Mathematical Sciences, University of Cambridge, Wilberforce Road, Cambridge CB3 0WA, UK}
%
\affiliation{Department of Physics and Astronomy, The University of
Mississippi, University, MS 38677, USA}
\affiliation{California Institute of Technology, Pasadena, CA 91109, USA}
\affiliation{Centro Multidisciplinar de Astrof{\'i}sica -- CENTRA, Departamento de F{\'i}sica, Instituto Superior T{\'e}cnico -- IST, 1049-001 Lisboa, Portugal}


\date{\today}

\begin{abstract}
We study the influence of astrophysical formation scenarios on the
precessional dynamics of spinning black-hole binaries by the time they
enter the observational window of second- and third-generation
gravitational-wave detectors, such as Advanced LIGO/Virgo, LIGO-India,
KAGRA and the Einstein Telescope. Under the plausible assumption that
tidal interactions are efficient at aligning the spins of few-solar
mass black-hole progenitors with the orbital angular momentum, we find
that black-hole spins should be expected to preferentially lie in a
plane when they become detectable by gravitational-wave
interferometers.  This ``resonant plane'' is identified by the
conditions $\Delta\Phi=0^\circ$ or $\Delta\Phi=\pm 180^\circ$, where
$\Delta\Phi$ is the angle between the components of the black-hole
spins in the plane orthogonal to the orbital angular momentum.  If the
angles $\Delta \Phi$ can be accurately measured for a large sample of
gravitational-wave detections, their distribution will constrain
models of compact binary formation.  In particular, it will tell us
whether tidal interactions are efficient and whether a mechanism such
as mass transfer, stellar winds, or supernovae can induce a mass-ratio
reversal (so that the heavier black hole is produced by the initially
lighter stellar progenitor). Therefore our model offers a concrete
observational link between gravitational-wave measurements and
astrophysics. We also hope that it will stimulate further studies of
precessional dynamics, gravitational-wave template placement and
parameter estimation for binaries locked in the resonant plane.
\end{abstract}

\maketitle 


\section{Introduction}

The inspiral and merger of stellar-mass black-hole (BH) binaries is
one of the main targets of the future network of second-generation
gravitational-wave (GW) interferometers (including Advanced LIGO/Virgo
\cite{2010CQGra..27h4006H}, LIGO-India \cite{indigo} and KAGRA
\cite{Somiya:2011np}) and of third-generation interferometers, such as
the proposed Einstein Telescope \cite{2010CQGra..27s4002P}.  Typical
GW signals from these binaries are expected to have low
signal-to-noise ratios, and must therefore be extracted by matched
filtering, which consists of computing the cross-correlation between
the noisy detector output and a predicted theoretical waveform, or
template (see e.g.~\cite{Sathyaprakash:2009xs}).  The number of
observationally distinguishable merger signals should be extremely
large, both because of the large and strongly mass-dependent number of
cycles in each signal and because the emitted waveform depends
sensitively on as many as 17 different parameters, in the general case
where the BHs are spinning and in eccentric orbits. The difficult
task of exploring such a high-dimensional space can be simplified if
nature provides physical mechanisms that cause astrophysical binaries
to cluster in restricted portions of the parameter space.

\begin{figure*}[thb]
\includegraphics[width=0.9\textwidth]{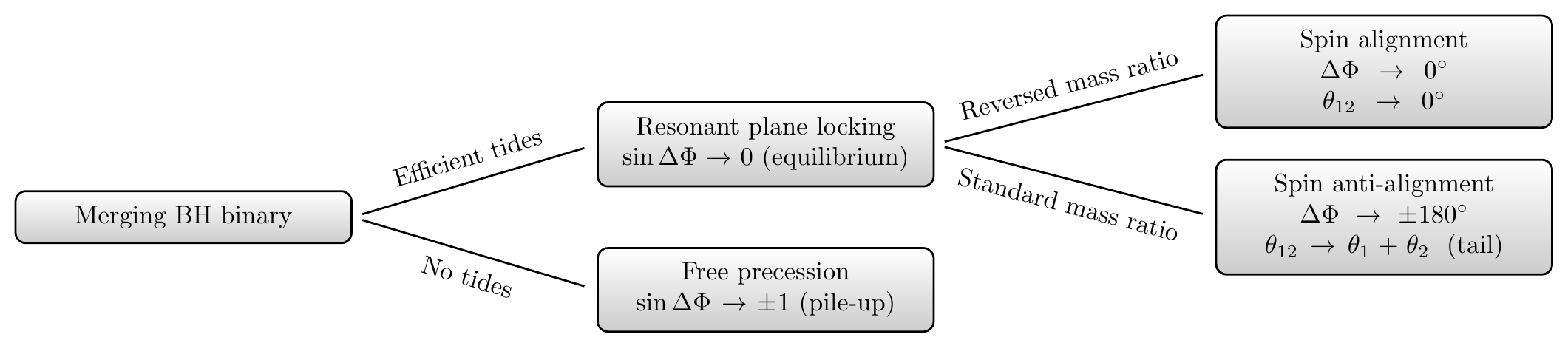}
\caption{Schematic summary of our predictions for the spin orientation
  of BH binaries as they enter the LIGO/Virgo band.  If tides
  efficiently align the spin of the secondary with the orbital angular
  momentum prior to the second supernova, resonant-plane locking will drive
  $\sin \Delta \Phi \to 0$, while in the absence of tides the spins
  will precess freely, piling up around $\sin \Delta \Phi \to \pm 1$
  near merger.  When tides are efficient, if the primary star evolves
  into the less massive BH (reversed mass ratio) the PN evolution will
  drive $\Delta \Phi \to 0^\circ, \theta_{12} \to 0^\circ$.  If
  instead the primary star evolves into the more massive BH (standard
  mass ratio) the PN evolution will drive $\Delta \Phi \to
  \pm180^\circ, \theta_{12} \to \theta_1 + \theta_2$, generating a
  tail in the distribution of $ \theta_{12}$ to larger values.  See
  Eqs.~(\ref{theta1theta2}) and (\ref{dphitheta12}) for definitions of
  these angles.}
\label{scheme}
\end{figure*}

In this paper we consider one mechanism to preferentially populate
certain regions of parameter space: the post-Newtonian (PN) spin-orbit
resonances first discovered by Schnittman \cite{2004PhRvD..70l4020S}.
Unfortunately, very few of the existing population-synthesis models of
compact-binary formation (see
e.g.~\cite{2008ApJ...682..474B,2010ApJ...719L..79F}) include
self-consistent predictions for BH spins.  To highlight the
significance of spin-orbit misalignment and resonances, we adopt a
simplified model for binary BH formation.  We use this model to
generate initial conditions for our compact binaries, and then
integrate the PN equations of motion forward in time using an
extension of the code used by some of us in previous studies of
supermassive BH binaries
\cite{2010PhRvD..81h4054K,2010ApJ...715.1006K,2012PhRvD..85l4049B}.
Our analytically tractable model captures (at least qualitatively)
many of the detailed physical effects influencing the evolution of BH
spins. Within this framework we carry out Monte Carlo simulations to
study the statistical distribution of BH spins when they enter the
GW-detection band of second- and third-generation detectors.

Before summarizing our results, we first introduce some notation. Consider
a BH binary with component masses $m_1 \geq m_2$, total mass $M=m_1+m_2$
and mass ratio $q=m_2/m_1 \leq 1$. 
The spin $\mathbf{S_i}$ of each BH can be written as
\begin{align}
\mathbf{S_i}=\chi_i \frac{Gm_i^2}{c} \mathbf{\hat S_i}
\,,
\end{align}
where $0 \leq \chi_i \leq 1$ ($i=1,2$) is the dimensionless spin
magnitude and a hat denotes a unit vector.
Our goal is not to rival the complexity of existing population-synthesis
models of compact-binary formation, but rather to
investigate specifically those astrophysical ingredients which
affect the spin dynamics.  We therefore focus on
maximally spinning BH binaries with mass ratio $q=0.8$, a typical
value predicted by population-synthesis studies (cf. e.g. Fig.~9
of \cite{2012ApJ...759...52D}).

Let us define $\theta_i$ to be the angle between each spin
$\mathbf{S_i}$ and the orbital angular momentum of the binary
$\mathbf{L}$, $\theta_{12}$ to be the angle between $\mathbf{S_1}$ and
$\mathbf{S_2}$, and $\Delta\Phi$ to be the angle between the
projection of the spins on the orbital plane:
\begin{alignat}{2}
\cos\theta_1&= \mathbf{ \hat  S_1}\cdot \mathbf{ \hat L}, &\quad \;\, 
 \cos\theta_2&=\mathbf{ \hat S_2}\cdot \mathbf{\hat L}, 
 \label{theta1theta2}
 \\
\cos\theta_{12}&=\mathbf{\hat S_1}\cdot \mathbf{\hat S_2}  ,  &\quad
\cos\Delta\Phi&=
\frac{ \mathbf{\hat S_1} \times \mathbf{\hat L} }
{| \mathbf{\hat S_1} \times \mathbf{\hat L}|}
\cdot
\frac{ \mathbf{\hat S_2} \times \mathbf{\hat L} }
{| \mathbf{\hat S_2} \times \mathbf{\hat L}|}.
\label{dphitheta12}
\end{alignat}
As we will demonstrate below, the physical mechanisms leading to the
formation of the BH binary leave a characteristic imprint on the
angles $\Delta \Phi$ and $\theta_{12}$. This has implications for GW
data analysis and, even more strikingly, for GW astronomy: at least in
principle, measurements of spin orientation with future GW detections
can constrain the astrophysical evolutionary processes that lead the
binary to merger.

All BH binaries with misaligned spins ($\theta_i \neq 0$) experience
PN spin precession as they inspiral towards merger.  Although
ensembles of BH binaries with isotropic spin distributions retain
their isotropic distributions as they inspiral
\cite{2007ApJ...661L.147B}, anisotropic spin distributions can be
substantially affected by PN spin precession
\cite{2004PhRvD..70l4020S}.  In particular, binaries can be attracted
towards PN spin-orbit resonances in which the BH spins and orbital
angular momentum jointly precess in a common plane (``resonant-plane
locking'').  Binaries in which the two BH spins and the orbital
angular momentum do not share a common plane at the end of the
inspiral are said to precess freely.  Binaries can become locked into
resonance if they satisfy the following conditions at large
separations:

\begin{itemize}

\item[i)] comparable but not equal masses ($0.4 \lesssim q \neq 1$),
\item[ii)] sufficiently large spin magnitudes ($\chi_i \gtrsim 0.5$),
\item[iii)] unequal spin misalignments ($\theta_1 \neq \theta_2$).
 
\end{itemize}

If these conditions are satisfied, the spin distribution of an
ensemble of binaries will be strongly influenced by the PN resonances
although every individual member of the ensemble will not necessarily
become locked into resonance.  In ensembles of binaries for which
$\theta_1 < \theta_2$ at large separations, the two spins tend to
align with each other, so that $\Delta\Phi \to 0^\circ$,
$\theta_{12}\to 0^\circ$.  If instead $\theta_1 > \theta_2$, the
projections of the BH spins on the orbital plane tend to anti-align,
so that $\Delta \Phi \to 180^\circ, \theta_{12} \to \theta_1 +
\theta_2$.  The mass ratios for which resonant-plane locking is
effective, given by condition i) above, are typical for the
stellar-mass BH binaries detectable by Advanced LIGO/Virgo (cf. Fig.~9
of \cite{2012ApJ...759...52D}).  The spin magnitudes $\chi_i$ of newly
formed BHs are highly uncertain, but observations of accreting BHs in
binary systems indicate that their spins span the whole range $0 \leq
\chi_i \leq 1$ allowed by general relativity
\cite{2009CQGra..26p3001B,2013ApJ...762..104S}.  Many BH-BH systems
may therefore satisfy condition ii) above.  In contrast, we would not
expect resonance locking in binaries in which one or both members are
neutron stars, as they are expected to have small
spins\footnote{Relativistic calculations of neutron star structure
  suggest that $\chi_i\lesssim 0.7$ for uniform rotation and
  physically motivated equations of state
  \cite{Cook:1993qr,Berti:2003nb,Lo:2010bj}, but the spin magnitudes
  of neutron stars in binaries observable by Advanced LIGO are likely
  to be much smaller than this theoretical upper bound
  \cite{Mandel:2009nx,Brown:2012qf}. The spin period of isolated
  neutron stars at birth should be in the range 10-140~ms
  \cite{lrr-2001-5}, or $\chi_i\lesssim 0.04$. Accretion from a binary
  companion can spin up neutron stars but is unlikely to produce
  periods less than 1 ms, i.e. $\chi_i\lesssim 0.4$
  \cite{2008AIPC.1068...67C}. The fastest-spinning observed pulsar has
  a period of 1.4~ms, ($\chi_i\sim 0.3$) \cite{Hessels:2006ze}; the
  fastest known pulsar in a neutron star-neutron star system,
  J0737-3039A, has a period of 22.70~ms ($\chi_i \sim 0.02$)
  \cite{Burgay:2003jj}.}.  Whether the spin misalignments of the
ensemble of BH binaries detectable by Advanced LIGO/Virgo is
asymmetric (satisfying condition iii) above) is a primary
consideration of this paper.

Astrophysical formation channels determine the initial conditions for
PN evolutions in the late inspiral. As a result they determine whether
resonant locking can occur, and which resonant configuration is
favored. Here we introduce a model for BH binary formation that allows
us to establish a link between binary-formation channels and the
near-merger spin configurations of precessing BH binaries. 

\begin{figure*}
\includegraphics[width=0.485 \textwidth]{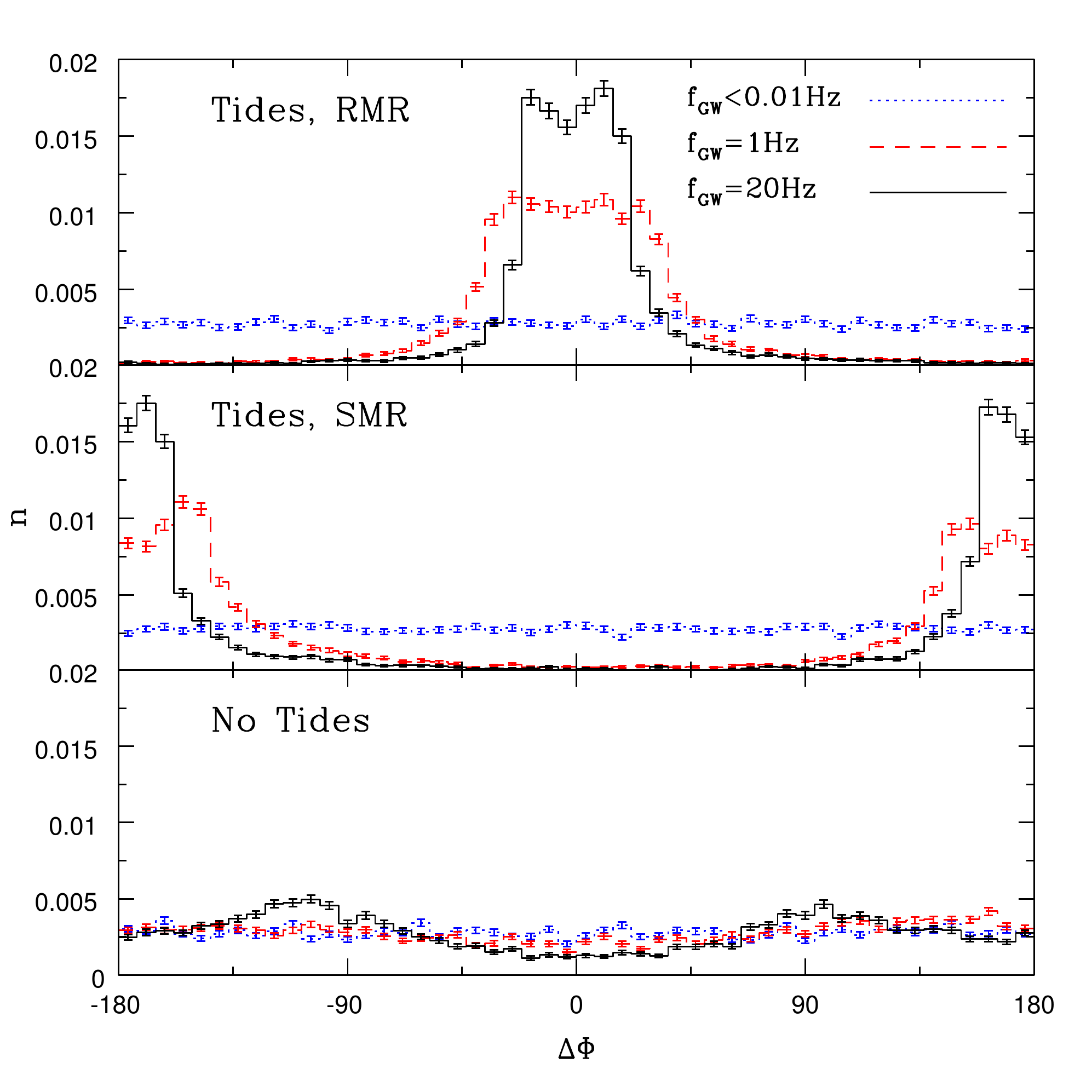}
\includegraphics[width=0.485 \textwidth]{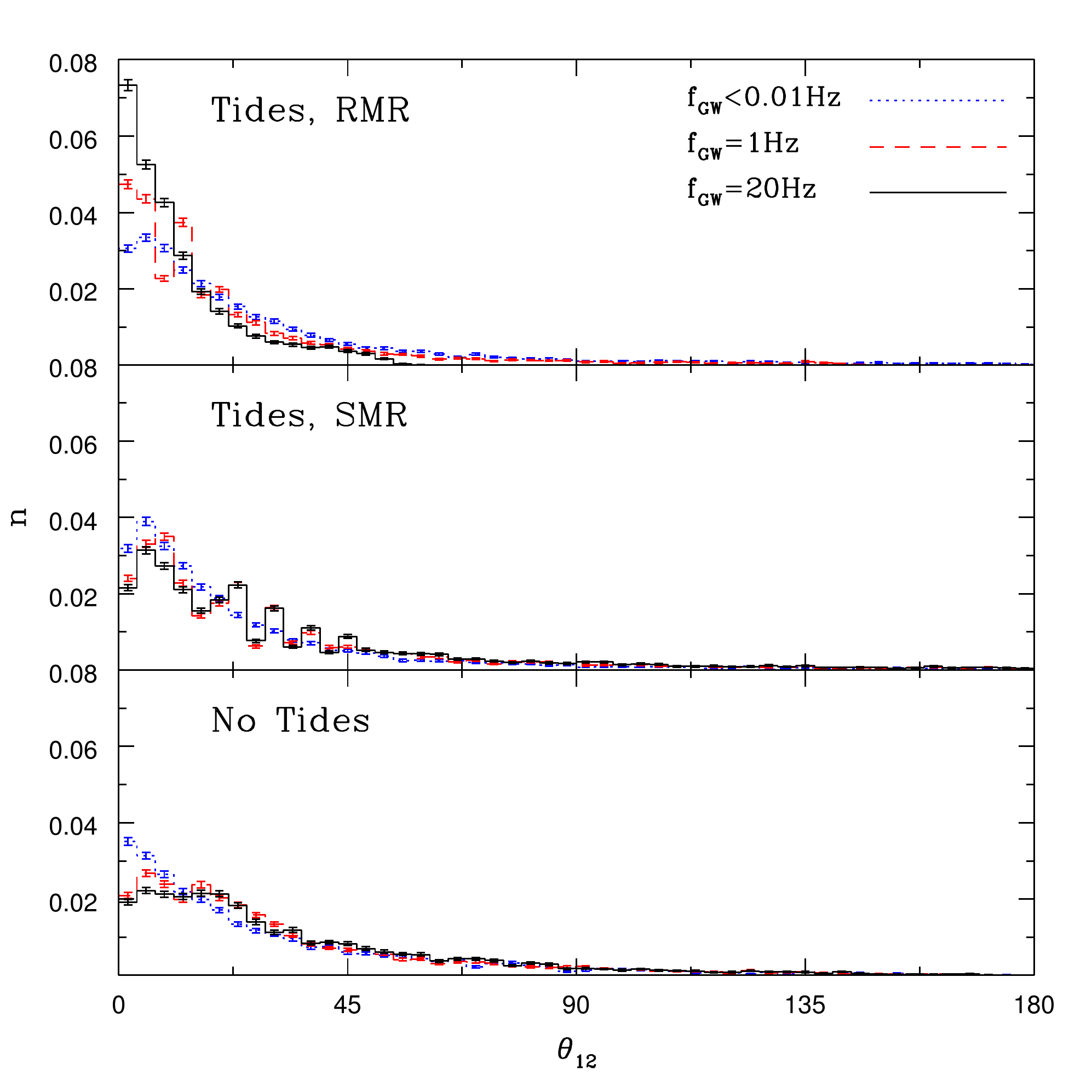}
\caption{(Color online.) Left: Probability distribution of the angle
  between the projections of the spins on the orbital plane
  $\Delta\Phi$.  As the binaries inspiral, the GW frequency $f_{\rm
    GW}$ increases from 0.01~Hz (dotted blue lines) to 1~Hz (dashed
  red lines) and later 20~Hz (solid black lines).  Under the effect of
  tides the PN evolution brings the spins in the same plane
  ($\Delta\Phi \to 0^\circ, \pm 180^\circ$), both in a reversed mass
  ratio (RMR, top panel) and in a standard mass ratio (SMR, middle
  panel) scenario. When tidal effects are removed (bottom panel, where
  we show both RMR and SMR binaries) the spins precess freely and pile
  up at $\Delta\Phi=\pm 90^\circ$. Right: Probability distribution of
  the angle between the two spins $\theta_{12}$. In the RMR scenario
  (top panel) the spins end up almost completely aligned with each
  other, i.e. most binaries have $\theta_{12}\simeq 0^\circ$. In the
  SMR scenario (middle panel) and in the absence of tides (bottom
  panel, where again we show both RMR and SMR binaries) a long tail at
  large values of $\theta_{12}$ remains even in the late inspiral. All
  simulations shown in this figure assume that kick directions are
  isotropically distributed. Error bars are computed assuming
  statistical Poisson noise.}
\label{intro_phi}
\end{figure*}

\subsection{Executive summary}

Our main findings are summarized schematically in
Fig.~\ref{scheme}. Supernova (SN) kicks tilt the orbit, producing a
misalignment between the orbital angular momentum and the orientation
of the spins of the binary members \cite{2000ApJ...541..319K}. As a
result, the main factors determining the spin alignment of a BH binary
are the magnitude of SN kicks and the possibility that other physical
effects may realign the spins with the orbital angular momentum in
between SN events. Dominant among these physical effects (aside from
the SN kick itself) are the efficiency of tidal interactions and the
possibility of a mass-ratio reversal due to mass transfer from the
initially more massive, faster evolving progenitor.

Tides affect the binary in two significant ways: they align the spins
of stellar BH progenitors with the orbital angular momentum and they
reduce the binary eccentricity. Additionally, tides force stars to
rotate synchronously with the orbit, increasing the likelihood of a
large BH spin at collapse and implying that our results will depend
only mildly (if at all) on the initial stellar spin. Consider the
evolution of the system between the two SN events, when the binary
consists of a BH and a non-degenerate star.  If tidal interactions are
efficient (a reasonable assumption, as we argue in
Appendix~\ref{tides}) they tend to align the star (but not the BH)
with the orbital angular momentum.  This introduces an asymmetry in
the angles $(\theta_1,\theta_2)$ which is critical to determining the
spin configuration at the end of the inspiral.

Mass transfer can change the mass ratio of interacting binaries.
Since the main-sequence lifetime of a star is a decreasing
function of its mass, the initially more massive star in a binary
is expected to collapse first.  If mass transfer from this star to its
less massive companion is insufficient, which we will refer to as the
{\em standard mass ratio} (SMR) scenario, the initially more massive
star will go on to form the more massive member of the BH binary.  
We cannot however rule out the possibility that prior to the first SN,
the initially more massive star overflows its Roche lobe and donates
mass to its initially lighter, longer-lived companion.  This mass
transfer may produce a mass-ratio reversal, so that the heavier BH in
the binary forms second: we will call this the {\em reversed mass
  ratio} (RMR) scenario.  According to population-synthesis models,
mass-ratio reversal happens for a sizable fraction (typically from
$\sim 10\%$ to $50\%$) of the total number of BH binaries
(cf.~\cite{2012ApJ...759...52D} and Table \ref{dominiktable} below).

Since BHs are relatively immune to the effects of tides, the spin of the
first BH to form will be more misaligned than the spin of the second
BH, as this misalignment will have accumulated due to the kicks generated
during {\it both} SN events.  Therefore, in the SMR scenario BH
binaries will have $\theta_1>\theta_2$ at formation, and thus
$\Delta \Phi\simeq \pm 180^\circ$ by the time they enter the
GW-detection band.
On the other hand, in the RMR scenario BH binaries initially have
$\theta_1<\theta_2$, so that by late in the inspiral $\Delta
\Phi\simeq \pm 0^\circ$, and furthermore the spins are nearly aligned
with each other (i.e., $\theta_{12}\simeq 0$).
In summary, whenever tidal interactions are efficient, our model
predicts that BH spins should preferentially lie in a ``resonant
plane'' (identified by the conditions $\Delta\Phi=0^\circ$ in the RMR
scenario, and $\Delta\Phi=\pm 180^\circ$ in the SMR scenario) when
they become detectable by GW interferometers.

A third (more unlikely) possibility is that tidal interactions are not
efficient. In this case, binaries form with $\theta_1 \simeq \theta_2$
and will not become locked into resonant configurations. Our
simulations show that binaries will preferentially have $\Delta
\Phi\simeq \pm 90^\circ$. Because the most likely values of $\Delta
\Phi$ in the three scenarios (RMR, SMR and no tides) are mutually
exclusive, GW measurements of a statistically significant sample of
values of $\Delta \Phi$ will provide important astrophysical
information on compact-binary formation scenarios. In particular, they
will tell us whether tidal interactions are efficient, and (if so)
whether mass transfer can produce mass-ratio reversals.

Fig.~\ref{intro_phi} makes these conclusions more quantitative by
showing three histograms of $\Delta\Phi$ (left) and $\theta_{12}$
(right), corresponding to snapshots taken at different times during
the inspiral. The distribution of $\Delta\Phi$ is flat at large
separations (dotted lines, corresponding to early times and small
orbital frequency) because spin-spin couplings are weak, and the BH
spins simply precess about the orbital angular momentum.  If tidal
alignment is efficient, in the late inspiral the BH spins lock into
equilibrium configurations with either $\Delta\Phi=0^\circ$ or
$\Delta\Phi=\pm 180 ^\circ$. This effect is clearly visible at GW
frequencies $f_{\rm GW}=1$~Hz, roughly corresponding to the lowest
cutoff frequency of third-generation detectors like ET, and it is even
more pronounced when the binaries enter the Advanced LIGO/Virgo band
at $f_{\rm GW}\simeq 20$~Hz. If tides are artificially removed, free
precession during the late stages of the inspiral slows down the
evolution of $\Delta\Phi$ when the components of the spin orthogonal
to the orbital angular momentum are also orthogonal to each other,
causing binaries that are not locked into resonance to pile up at
$\Delta\Phi=\pm 90^\circ$.

Let us stress again that the statistical effect of resonances is
clearly visible at $f_{\rm GW}=20 {\rm \,Hz}$, i.e. when BH binaries
enter the Advanced LIGO/Virgo band.  GW measurements of $\Delta\Phi$
can therefore be used to constrain uncertainties in BH
binary-formation scenarios.  The inclusion of resonant effects in
population-synthesis models (combined with a statistically significant
sample of GW measurements of $\Delta\Phi$) has the potential to
constrain various aspects of the models, such as the efficiency of
tides, stable mass transfer, common-envelope (CE) evolution, SN kick
velocities, and the metallicity of BH progenitors.

\subsection{Outline of the paper}

The rest of the paper provides details of our astrophysical model and
a more detailed discussion of the results.  In
Section~\ref{binformation} we introduce our fiducial BH
binary-formation channels, which are based on detailed
population-synthesis models, as described at much greater length in
Appendix~\ref{ap:BinEv}. In order to focus on spin effects, we fix the
component masses to two representative values.  We assume that SN
kicks follow a Maxwellian distribution in magnitude.
We also assume that the kicks are distributed in a double cone of
opening angle $\theta_b$ about the spin of the exploding star and, to
bracket uncertainties, we consider two extreme scenarios: isotropic
($\theta_b=90^\circ$) or polar ($\theta_b=10^\circ$) kicks.

Section~\ref{PNinspiral} summarizes the results of evolving these BH
binaries under the effect of gravitational radiation down to a final
separation of $10GM/c^2$. We demonstrate that spin-orbit resonances
have a significant impact on the observable properties of our fiducial
BH binaries.  Although we have only explored a handful of evolutionary
channels and component masses, in Section~\ref{sec:CompareToPopsyn} we
argue that the scenarios described in Fig.~\ref{scheme} are broadly
applicable: kicks, tides, and the mass-ratio distribution control spin
alignment.  We explore the sensitivity of these three features (and
hence of the observable distribution of resonantly-locked binaries) to
several poorly constrained physical inputs to binary-evolution models,
and we argue that GW observations of precession angles could provide
significant constraints on binary formation channels. Finally, in
Section~\ref{results} we describe the implications of our results for
future efforts in binary-evolution modeling and GW detection.

To complement and justify the simple astrophysical model proposed in
Section~\ref{binformation}, in Appendix~\ref{ap:BinEv} we describe in
detail the rationale underlying the model and its relationship to our
current understanding of binary evolution. Appendix~\ref{ap:BinEv}
should provide a useful resource to implement (and possibly improve)
the Monte-Carlo algorithm described in the main text.

\section{Astrophysical model of the initial conditions for spin evolution}
\label{binformation}

Isolated BH binaries do not emit electromagnetically and hence have
yet to be observed.  Despite this lack of evidence, they are a likely
outcome of the evolution of massive stellar binaries.  The rate at
which they form can be inferred from observations of their progenitors
and systems like binary neutron stars that have similar formation
channels.  Formation rates can also be calculated theoretically using
population-synthesis models such as \texttt{StarTrack}
\cite{2002ApJ...572..407B,2008ApJS..174..223B,2012ApJ...757...91B,2012ApJ...759...52D},
which builds upon previous analytical studies of single
\cite{2000MNRAS.315..543H} and binary stellar evolution
\cite{2002MNRAS.329..897H}.

Most studies of compact-binary formation do not keep track of the
magnitude and orientation of BH spins, and those that do (see
e.g.~\cite{OShaughnessy:2005qc,2008ApJ...682..474B,2010ApJ...719L..79F})
neglect general-relativistic effects in the late-time evolution of the
binary.  One of the goals of our study is to fill this gap.
For example, the version of the \url{StarTrack} code used in
\cite{2008ApJ...682..474B} assumed that both $\mathbf{S_1}$ and
$\mathbf{S_2}$ remained aligned with the initial direction of the
orbital angular momentum $\mathbf{L}$.  
The evolution of $\mathbf{L}$ itself was performed by applying energy and angular-momentum
conservation when compact objects are formed (and kicked) as a result
of gravitational collapse.  This approach is suitable for binaries in
nonrelativistic orbits, like observed X-ray binaries
\cite{OShaughnessy:2005qc,2010ApJ...719L..79F}, but it may not be
appropriate for merging binaries, that are interesting both as GW
sources and as progenitors of short gamma-ray bursts
\cite{2008ApJ...682..474B}. Since existing BH binary-formation models
preserve the mutual alignment of the spins with the initial direction
of $\mathbf{L}$, all BH-BH binaries are formed with
$\theta_1=\theta_2$.  Later models of mixed BH X-ray binaries do allow
for the possibility of asymmetric spin configurations via accretion
\cite{2010ApJ...719L..79F}, but to the best of our knowledge no such
studies have been published for the BH-BH case.  Since PN resonance
locking only occurs when $\theta_1\neq \theta_2$, its effects are
excluded by construction in the BH binary models available in the
literature.

Here we develop a slightly more complex (and presumably more
realistic) model for spin evolution, allowing for the formation of
``asymmetric'' BH binaries with $\theta_1\neq\theta_2$. The model is
not meant to rival the complexity of population-synthesis codes like
\texttt{StarTrack}. Our goal is rather to isolate the physical
ingredients that are specifically relevant to BH spin alignment.  The
model builds, when necessary (e.g. when computing the remnant masses
resulting from gravitational collapse as a function of the progenitor
masses, or in treating the CE phase) on results from
\texttt{StarTrack}, and in Section~\ref{sec:CompareToPopsyn} we
present a preliminary comparison of our conclusions with publicly
available results from \texttt{StarTrack}.

\subsection{Length scales}

Before describing our astrophysical model, we review the length scales
associated with the formation, inspiral, and merger of BH binaries.
The well defined hierarchy in these length scales demonstrates the
necessity of our joint analysis of astrophysics and PN evolution.  GW
emission \cite{1964PhRv..136.1224P,1963PhRv..131..435P} causes a
binary with a semimajor axis less than
\begin{align}
a_{\rm H} \sim 45 \left[ \frac{q}{(1+q)^2} \left(\frac{t_{\rm GW}}{10^{10}~{\rm yrs}}\right) \left(\frac{M}{10 M_\sun}\right)^3
\right]^{1/4} R_\odot
\end{align}
to merge on a timescale $t_{\rm GW}$ less than the Hubble time $t_{\rm
  H}\simeq10^{10} {\rm yrs}$.  The astrophysical processes described
in this Section, including mass transfer, SN
explosions\footnote{Throughout the paper we will loosely use the term
  ``supernova'' to indicate the core collapse of massive stars, even
  when such events are not luminous.} and CE evolution, are required
to shrink the binary down to separations smaller than $a_{\rm H}$.  GW
emission also circularizes the binary at separations comparable to
$a_{\rm H}$.  PN spin-orbit couplings become important at much smaller
separations
 \begin{align}
a_{\rm PNi}  \sim 10^3 \frac{G M}{c^2}\simeq 10^{-2} \left(\frac{M}{10 M_\sun}\right) R_\sun~,
\end{align}
below which they can lock binaries into resonant configurations with
well defined spin directions \cite{2004PhRvD..70l4020S}.  Previous
studies of PN resonances for supermassive BHs
\cite{2010PhRvD..81h4054K,2010ApJ...715.1006K,2012PhRvD..85l4049B}
found that the effectiveness of resonance locking strongly depends on
the orientation of the BH spins when the binary reaches the separation
$a_{\rm PNi}$.  The spin orientation is set by the binary's
astrophysical formation history.  Resonance locking can be important
even at separations above
\begin{align}
a_{\rm LIGO} \simeq 10^{-3} \left(\frac{M}{10 M_\sun}\right)^{1/3}   \left( \frac{f_{\rm GW} }{20 {\rm Hz}}\right)^{-2/3} R_\sun \,,
\label{sepfreq}
\end{align}
at which the binary reaches the lower limit $f_{\rm GW}\simeq 10-20
\,{\rm Hz}$ of the Advanced LIGO/Virgo sensitivity band.  The
third-generation Einstein Telescope is expected to reach even lower
frequencies of order $f_{\rm GW} \simeq1 \,\rm{Hz}$.  Since these
frequencies are well within the regime where PN resonances are
important, a unified treatment of the astrophysical initial conditions
and of the subsequent PN evolution of the binary is essential to
determining which spin configurations are most relevant for GW
detectors.  Such a treatment is the main goal of this work.

\begin{figure*}[htb]
\includegraphics[width=\textwidth]{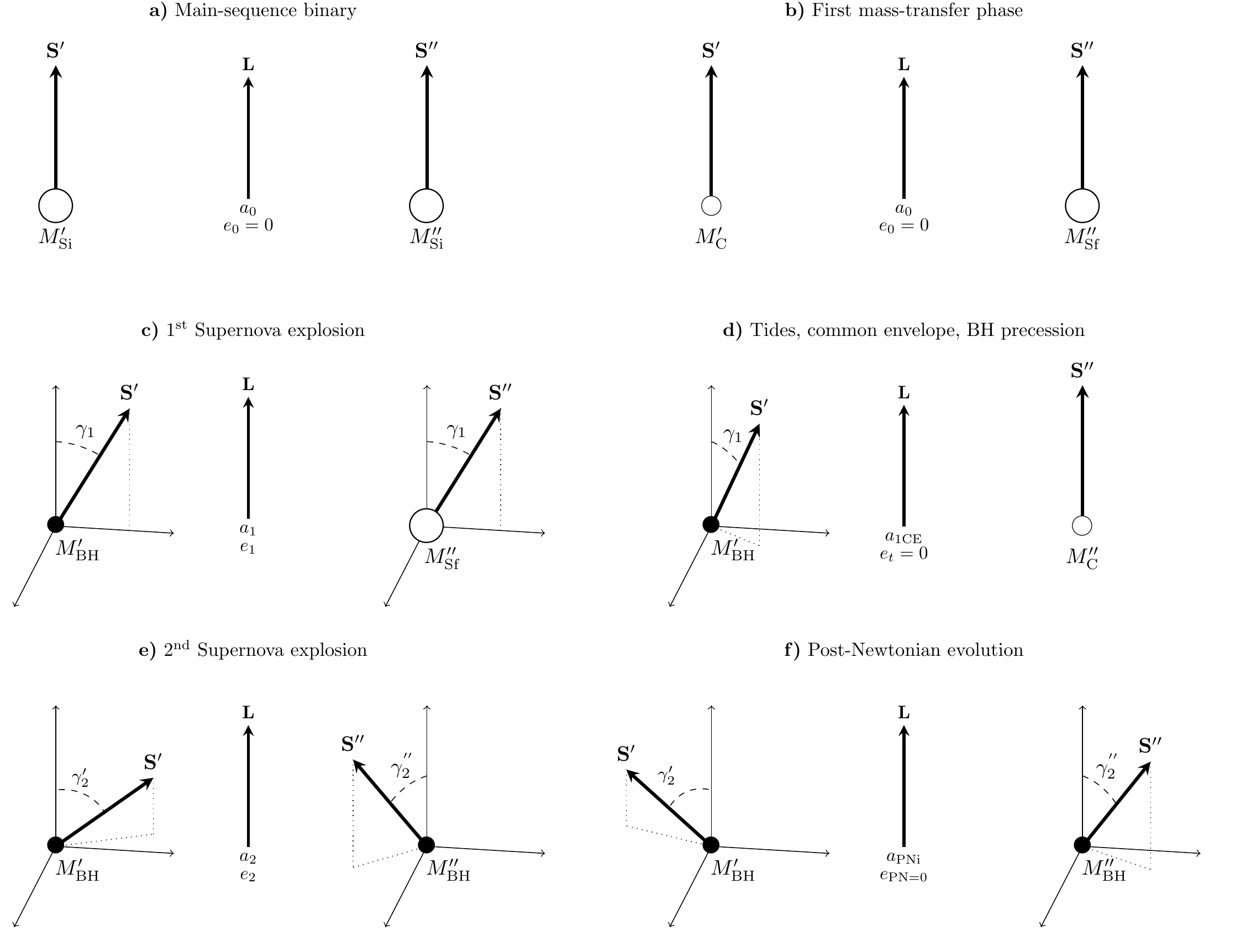}
\caption{A schematic representation of our model for BH binary
  formation and spin evolution. Empty circles represent stars, filled
  circles represent BHs.  Phase (\textbf{a}) shows the initial
  main-sequence stellar binary.  Mass transfer from the primary to the
  secondary (\textbf{b}) leads to a possible mass-ratio reversal.  The
  first SN kick tilts the angle between the spins and the orbital
  plane (\textbf{c}).  Tidal interactions can realign the stellar
  member of the binary (\textbf{d}).  The second SN kick tilts the
  orbital plane again (\textbf{e}).  Gravitational radiation shrinks
  and circularizes the binary before our explicit PN evolution begins
  (\textbf{f}).  }
\label{model}
\end{figure*}

\subsection{Fiducial scenarios for binary evolution} \label{SS:example}

In this Section we describe how massive main-sequence binary stars
evolve into BH binaries.  Fig.~\ref{model} summarizes the critical
stages of binary evolution in our model.  To isolate the effects of
spin orientation during the PN inspiral of the BH binaries, we fix the
final mass ratio to the typical value $q = 0.8$
\cite{2012ApJ...759...52D}.  To ensure that this final mass ratio is
obtained, the initial stellar masses of the binaries must be fixed to
$(M'_{Si},M''_{Si}) = (35 M_\odot, 16.75 M_\odot)$ in the SMR
scenario, or $(30 M_\odot,24 M_\odot$) in the RMR scenario.
Throughout the paper, we use a single prime to identify the initially
more massive stellar progenitor or ``primary'', and a double prime to
denote the initially less massive progenitor or ``secondary''.
This choice of initial masses also fixes the total mass of our BH
binaries to $M = 13.5 M_\odot$, quite close to the expected peak of
the distribution for the total mass \cite{2012ApJ...759...52D}.  The
mass of the stars is somewhat smaller than expected for the
progenitors of BHs of these masses because we have neglected stellar
winds, that lead to considerable mass loss prior to BH formation.
Table~\ref{numbers} provides numerical values for the masses and radii
of both the primary and secondary throughout the evolution in both the
SMR and RMR scenarios.  Appendix~\ref{A:InitMass} shows how this
choice of initial masses leads to BHs of the desired final masses.

\begin{table}[hbt]
\begin{center}
\begin{tabular}{c@{\hskip 7pt}r@{\hskip 7pt}r@{\hskip 25pt}c@{\hskip 7pt}r@{\hskip 7pt}r}
\hline \hline
& SMR $\;$& RMR $\;$& & SMR $\;$& RMR $\;$\\
\hline \hline
$ M'_{\rm Si}$ & $35 M_\odot$ & $30 M_\odot$  &			$ R'_{\rm Si}$ &  $9.57 R_\odot$ &   $8.78 R_\odot$\\	
$ M''_{\rm Si}$ & $16.75 M_\odot$ & $24 M_\odot$ &		$ R''_{\rm Si}$ &  $6.36 R_\odot$&  $7.76 R_\odot$\\	
$ M''_{\rm Sf}$ & $30 M_\odot$ & $35 M_\odot$ &			$ R''_{\rm Sf}$ &  $8.78 R_\odot$&  $9.57 R_\odot$\\
$ M'_{\rm C}$ & $8.5 M_\odot$ & $8 M_\odot$ &			$ R'_{\rm C}$ &  $0.26 R_\odot$&  $0.26 R_\odot$\\	
$ M''_{\rm C}$ & $8 M_\odot$ & $8.5 M_\odot$ &			$ R''_{\rm C}$ &  $0.27 R_\odot$&  $0.27 R_\odot$\\
$ M'_{\rm BH}$ & $7.5 M_\odot$ & $6 M_\odot$ &			$ R'_{\rm G}$ &  $3608 R_\odot$&  $3500 R_\odot$\\	
$ M''_{\rm BH}$ & $6 M_\odot$ & $7.5 M_\odot$ &			$ R''_{\rm G}$ &  $3500 R_\odot$&  $3608 R_\odot$\\

$ a_{\rm min}$ & $17.9  R_\odot$&  $18.8 R_\odot$&		$ a_{\rm noCE}$ & $6981  R_\odot$&  $6758 R_\odot$\\
$ a_{\rm max}$ &  $8128 R_\odot$&  $8787 R_\odot$&		$ a_{\rm mCE}$ &  $0.69 R_\odot$&  $0.63 R_\odot$\\
\hline\hline
\end{tabular}
\end{center}
\caption{Masses and length scales at various stages of the binary
  evolution in our SMR and RMR scenarios, as shown in
  Fig.~\ref{model}.  The only independent parameters are the
  main-sequence masses $M'_{\rm Si}$ and $M''_{\rm Si}$, which have
  been tuned to study final BH binaries with mass ratio $q=0.8$. The
  other values are defined in the main text, and they are obtained
  using the analytical prescriptions presented in
  Appendix~\ref{ap:BinEv}.}
\label{numbers}
\end{table}

The initial main-sequence stage of the evolution is shown as phase
\textbf{a} in Fig.~\ref{model}.  Binaries are assumed to form on
circular\footnote{The initial eccentricity has minimal effect. In fact
  we have repeated our calculations using an initially thermal
  distribution of eccentricities of the form $f(e)=2e$, and we
  observed no significant difference in the final distribution of
  $\Delta \Phi$ and $\theta_{12}$.} orbits with initial semimajor axes
$a_0$ drawn from the distribution described in
Appendix~\ref{A:InitSep}.  We assume that the spins of the primary
$\mathbf{S^\prime}$ and secondary $\mathbf{S^{\prime\prime}}$ are
initially aligned\footnote{The alignment of stellar spins in eclipsing
  binaries can be measured through the Rossiter-McLaughlin effect
  \cite{1924ApJ....60...15R,1924ApJ....60...22M}.  Although many
  systems have aligned spins
  \cite{2007A&A...474..565A,2011ApJ...726...68A,2012arXiv1211.7065A},
  there are notable exceptions \cite{2009Natur.461..373A}.  We expect
  efficient tidal alignment in the progenitors of merging BH binaries,
  due to their small initial separations.} with the orbital angular
momentum $\mathbf{L}$.  As the primary evolves, its envelope expands
until it fills its Roche lobe, initiating stable mass transfer to the
secondary (phase \textbf{b} in Fig.~\ref{model}). The efficiency of
mass transfer is usually parametrized via a parameter $f_a\in
[0,\,1]$: cf. Eq.~(\ref{fa_def}) of Appendix~\ref{A:MT}.  We assume
this mass transfer continues until the primary has depleted its
hydrogen envelope, leaving behind a helium core of mass $M_C'=8.5
M_\odot$ ($M_C'=8 M_\odot$) in the SMR (RMR) scenario. Following
\cite{2012ApJ...759...52D}, we assume semiconservative mass transfer:
the secondary accretes a fraction $f_a = 1/2$ of the mass lost by the
primary, growing to a mass $M_{Sf}''=30 M_\odot$ ($M_{Sf}''=35
M_\odot$) in the SMR (RMR) scenario at the end of the mass-transfer
episode. In principle mass transfer should also change the orbital
separation, but we neglect this change as it is smaller than the width
of the distribution of initial separations, as well as subsequent
changes in the separation during the CE phase.

Following the end of mass transfer, the primary explodes in a SN
(phase \textbf{c} in Fig.~\ref{model}) producing a BH of mass $M_{BH}'
= 7.5~M_\odot$ ($M_{BH}'=6~M_\odot$ ) in the SMR (RMR) scenario. For
simplicity, in our simulations the spin of this newly born BH is
assumed to be maximal ($\chi_i = 1$, $i=1\,,2$) and aligned with its
stellar progenitor. The SN ejecta are generally emitted
asymmetrically, imparting a recoil velocity to the BH which is
generally a fraction of the typical recoil velocities for protoneutron
stars: $v_{\rm BH} \simeq (1 - f_{\rm fb}) v_{\rm pNS}$, where $f_{\rm
  fb}\in [0,\,1]$ is a ``fallback parameter''
(cf. Appendix~\ref{A:kick}).  This recoil tilts the orbital plane by
an angle $\gamma_1$, and changes the semimajor axis and eccentricity
to $a_1$ and $e_1$, respectively.  These orbital changes depend on
both the kick and the mass lost during the SN, as described in
Appendix~\ref{SNkicks}.

After the SN explosion of the primary, the secondary evolves and
expands.  The primary raises tides on the swollen secondary, and
dissipation may allow these tides to both circularize the orbit (so
that the final eccentricity is $e_t \simeq 0$) and align the spin
$\mathbf{S^{\prime\prime}}$ of the secondary with the orbital angular
momentum $\mathbf{L}$, as shown in phase \textbf{d} of
Fig.~\ref{model}.  This tidal alignment is described in much greater
detail in Appendix~\ref{tides}.  Given the uncertainty in the
efficiency of tidal alignment, we explore both extreme possibilities:
complete circularization and alignment of $\mathbf{S^{\prime\prime}}$
(``Tides" in Fig.~\ref{intro_phi}) and no circularization and
alignment at all (``No Tides" in Fig.~\ref{intro_phi}).  As the
secondary expands further, it fills its Roche lobe initiating a second
phase of mass transfer.  However, unlike the first mass-transfer
event, this second mass-transfer phase will be highly unstable
\cite{1997A&A...327..620S,2010Ap&SS.329..243G,2012ApJ...746..186C}.
Instead of being accreted by the primary, most of this gas will expand
into a CE about both members of the binary.  Energy will be
transferred from the binary's orbit to the CE, ultimately unbinding it
from the system.  This energy loss shrinks the semimajor axis of the
binary from $a_1$ to $a_{\rm 1CE}$, as shown in phase \textbf{d} of
Fig.~\ref{model}.  More details about CE evolution, including the
relationship between $a_1$ and $a_{\rm 1CE}$, are provided in
Appendix~\ref{A:CE}.  After the secondary loses its hydrogen envelope,
the remaining helium core has a mass $M''_C = 8 M_\odot$ ($M''_C = 8.5
M_\odot$) in the SMR (RMR) scenario, as listed in Table~\ref{numbers}.

After the end of CE evolution, the naked helium core of the secondary
rapidly completes its stellar evolution and explodes as a SN, as shown
in phase \textbf{e} of Fig.~\ref{model}.  This explosion produces a BH
of mass $M''_{\rm BH} = 6 M_\odot$ ($M''_{\rm BH} = 7.5 M_\odot$) in
the SMR (RMR) scenario, as listed in Table~\ref{numbers}.  We assume
that this BH has a maximal spin that is aligned with the spin
$\mathbf{S^{\prime\prime}}$ of its stellar progenitor, as we did for
the primary.  The SN leads to mass loss and a hydrodynamical recoil
that change the semimajor axis and eccentricity of the binary to $a_2$
and $e_2$, respectively.  It also tilts the orbital plane by an angle
$\Theta$ that can be calculated using the same procedure as given for
the first SN in Appendix~\ref{SNkicks}. The tilt resulting from the
second SN is generally much smaller than that from the first SN
($\Theta \ll \gamma_1$) due to the comparatively larger orbital
velocity following CE evolution.  This tilt changes the angles between
$\mathbf{L}$ and the spins $\mathbf{S^{\prime}}$ and
$\mathbf{S^{\prime\prime}}$ to $\gamma^{\prime}_2$ and
$\gamma^{\prime\prime}_2$, respectively.  If tides efficiently align
$\mathbf{S^{\prime\prime}}$ with $\mathbf{L}$ prior to the second SN,
these angles are given by
\begin{align}
\cos \gamma'_2 &= \cos \gamma_1 \cos \Theta + 
\cos \varphi' \sin \gamma_1 \sin \Theta\,, 
\label{thetapT} \\
\cos \gamma''_2 &= \cos \Theta~\,, \quad ({\rm tides})
\label{thetappT}
\end{align}
where $\varphi^\prime$ is the angle between the projection of
$\mathbf{S^{\prime}}$ in the orbital plane before the SN and the
projection of the change in $\mathbf{L}$ into this same initial
orbital plane.  If $\varphi^\prime$ is uniformly distributed (the
direction of the SN kick of the secondary is uncorrelated with the
spin of the primary), the second term on the right-hand side of
Eq.~(\ref{thetapT}) averages to zero, implying that $\gamma'_2 >
\gamma''_2$ for most binaries\footnote{Well separated distributions of
  $\gamma'_2$ and $\gamma''_2$ require SN kick velocities that are
  comparable to the orbital velocity prior to the first SN, but much
  smaller than the orbital velocity before the second SN.  Fortunately
  such kick velocities are well motivated, as described in Appendix
  \ref{A:kick}.}.  This is the mechanism for creating a binary BH
population preferentially attracted to the $\Delta \Phi =
\pm180^\circ$ family of spin-orbit resonances in the SMR scenario and
the $\Delta \Phi = 0^\circ$ family of resonances in the RMR scenario,
as shown in Fig.~\ref{intro_phi}.

If tides are inefficient, $\gamma^{\prime\prime}_2$ is instead given by
\begin{align}
\cos \gamma^{\prime\prime}_2
&= \cos \xi~\, \quad ({\rm no~tides}) \\
&= \cos \gamma_1 \cos \Theta - 
\sin \varpi \sin \gamma_1 \sin \Theta\,, \label{thetappNT}
\end{align}
where $\xi$ is given by Eq.~(\ref{spintiltangle}), and $\varpi$ is the
angle between the projection of $\mathbf{S^{\prime\prime}}$ into the
orbital plane before the second SN and the separation vector between
the members of the binary.  If $\varpi$ is independent of
$\varphi^\prime$ and uniformly distributed\footnote{This assumption is
  well justified because the primary and secondary spins precess at
  different rates [$\Omega_1$ and $\Omega_2$ given by
    Eqs.~(\ref{precession1}) and (\ref{precession2}) below] and the
  precession timescale $t_{\rm pre} \sim \Omega_i^{-1}$ is short
  compared to the time $t_{\rm SN} \sim 10^6$~yrs between SN events.
  At lowest PN order, $t_{\rm pre} \sim t_{\rm LC} (v/c)^{-5}$, where
  $t_{LC} = GM/c^3\simeq 5\times 10^{-5}(M/10 M_\odot)$~s is the
  light-crossing time.  At a separation $a$ we have $v/c\sim 5\times
  10^{-3} (M/10M_\odot)^{1/2} (a/R_\odot)^{-1/2}$, so $t_{\rm pre}
  \sim 0.5~{\rm yr} \ll t_{\rm SN}$.}, the second term on the
right-hand side of Eq.~(\ref{thetappNT}) also averages to zero,
implying that $\gamma'_2 \simeq \gamma''_2$. The small scatter about
this relation follows from the lesser influence of the second SN kick
($\Theta \ll \gamma_1$), which implies that the identical first terms
on the right-hand sides of Eqs.~(\ref{thetapT}) and (\ref{thetappNT})
dominate over the differing second terms. This explains the lack of
preference for either family of resonances in the ``No Tides''
scenario shown in Fig.~\ref{intro_phi}.

After the second SN, the BH binary is left in a non-relativistic orbit
that gradually decays through the emission of gravitational radiation,
as shown in phase \textbf{f} of Fig.~\ref{model}.  We calculate how
this orbital decay reduces the semimajor axis and eccentricity using
the standard quadrupole formula
\cite{1964PhRv..136.1224P,Peters:1963ux}:
\begin{align}
\frac{dt}{da} &= -\frac{5}{64}\frac{c^5 a^3}{G^3 M^3 }    \frac{(1+q)^2}{q} 
(1-e^2)^{7/2} \left(1+\frac{73}{24}e^2+\frac{37}{96}e^4\right)^{-1},
 \label{petersdtda}\\
\frac{de}{da} &= \frac{19}{12}\frac{e}{a}   
(1-e^2)\left(1+\frac{121}{304} e^2\right)
\left(1+\frac{73}{24}e^2+\frac{37}{96}e^4\right)^{-1}.
\label{petersdeda}
\end{align}
To an excellent approximation, the BH spins simply precess about
$\mathbf{L}$ during this stage of the evolution, leaving
$\gamma^{\prime}_2$ and $\gamma^{\prime\prime}_2$ fixed to their
values after the second SN.  Once the semimajor axis reaches a value
$a_{\rm PNi} = 1000M$ (in units where $G = c = 1$), we integrate
higher-order PN equations of motion as described in
Section~\ref{PNinspiral} to carefully model how the orbit and spins
evolve.  We assume that radiation reaction circularizes the orbit
($e_{\rm PN} = 0$) by the time we start integrating the higher-order
PN equations describing the precessional dynamics of the BH
binary. This assumption is fully justified, as we will show by
explicit integration in Section \ref{PNinit} below.

\subsection{Synthetic black-hole binary populations}
\label{sec:sub:RandomEvolve}

\begin{table*}[thb]
\begin{tabular}{c@{\hskip 5pt}c@{\hskip 5pt}c@{\hskip 15pt}ll@{\hskip 7pt}ll@{\hskip 7pt}ll@{\hskip 7pt}ll@{\hskip 7pt}ll}
\hline \hline
Kicks &Tides &Mass transfer & \multicolumn{2}{c}{$\nu_{\rm SN1}(\%)$} &\multicolumn{2}{c}{$\nu_{\rm mCE}(\%)$} &\multicolumn{2}{c}{$\nu_{\rm SN2}(\%)$} &\multicolumn{2}{c}{$\nu_{\rm H}(\%)$} &\multicolumn{2}{c}{$\nu_{\rm BH}(\%)$}\\
\hline 
\hline
Isotropic     &On  &SMR &		 32.50 &(80.50) & 26.53 &(12.24) & 2.66 &(0.51) & 0.04 &(0.00) & 38.27 &(6.74) \\
Isotropic	 &On  &RMR & 		32.55 &(80.28) & 34.86 &(14.91) &  2.97  &(0.30) &  0.04  &(0.00) &  29.59 &(4.50) \\
\hline						
Isotropic &  Off & SMR & 		 32.50 &(80.50) & 26.53 &(12.24) &  2.93  &(0.60) &  0.04  &(0.01) &  38.01 &(6.65) \\
Isotropic &  Off & RMR & 		 32.55 &(80.28) &  34.86 &(14.91) &  3.01  &(0.35) &  0.04  &(0.00) &  29.54 &(4.46) \\
\hline						
Polar          &On  &SMR &		 31.84 &(83.14) &  26.68 &(9.40) &  3.29  &(0.24) &  0.01  &(0.01) &  38.18 &(7.21) \\
Polar	         &On  &RMR &		 31.86 &(82.97) &  34.88 &(12.10) &  3.65  &(0.24) &  0.02  &(0.00) &  29.58 &(4.70) \\
\hline						
Polar  &   Off   & SMR  & 		 31.81 &(83.16) &  26.65 &(9.38) &  3.35  &(0.52) &  0.03  &(0.01) &  38.15 &(6.93) \\
Polar   &  Off &   RMR  & 		 31.84 &(82.98) &  34.89 &(12.09) &  3.65  &(0.33) &  0.04  &(0.00) &  29.59 &(4.60) \\
\hline \hline
\end{tabular}
\caption{Fraction of binaries $\nu$ (in percentage) that satisfy the
  following conditions, each of which successively \emph{prevent} the
  formation of a merging BH binary: i) are unbound by the first SN
  ($\nu_{\rm SN1}$), ii) merge during the CE phase ($\nu_{\rm mCE}$),
  iii) are unbound by the second SN ($\nu_{\rm SN2}$), iv) do not
  merge within a Hubble time due to gravitational-radiation reaction
  ($\nu_{\rm H}$).  The final column is the fraction $\nu_{\rm BH}
  =1-(\nu_{\rm SN1}+\nu_{\rm mCE} + \nu_{\rm SN2} + \nu_{\rm H})$ of
  all simulated binaries that form merging BH binaries.  In
  parentheses we list the corresponding fractions if SN kicks are not
  suppressed by fallback (i.e. if we set $f_{\rm fb} = 0$ rather than
  $f_{\rm fb} = 0.8$): see Appendix~\ref{A:kick}).}
\label{rates}
\end{table*}

In the previous Section, we presented fiducial scenarios for the
formation of BH binaries characterized by three choices:

\begin{itemize}

\item[i)] stable mass transfer prior to the first SN can preserve
  (SMR) or reverse (RMR) the mass ratio of the binary;
\item[ii)] hydrodynamic kicks generated by the SN can have a polar
  ($\theta_b = 10^\circ$) or isotropic ($\theta_b = 90^\circ$)
  distribution with respect to the exploding star's spin;
\item[iii)] tides do or do not circularize the orbit and align the
  spin $\mathbf{S^{\prime\prime}}$ of the secondary with the orbital
  angular momentum $\mathbf{L}$ prior to the second SN.
 
\end{itemize}

In this Section, we construct synthetic populations of BH binaries for
the 8 different scenarios determined by the three binary choices
listed above.  To generate members of these synthetic populations, we
perform Monte Carlo simulations\footnote{We generated $10^8$ binary
  progenitors to calculate the rates listed in Table~\ref{rates},
  which are therefore accurate to within $\sim 0.01\%$. To avoid
  cluttering, we only show a subsample of $10^4$ progenitors in the
  figures of this Section.} in which random values determine

\begin{itemize}

\item[i)] the initial semimajor axis $a_0$
  (Appendix~\ref{A:InitSep}),
\item[ii)] the magnitude and direction of the kick produced in the
  first SN (Appendix~\ref{A:kick}),
\item[iii)] the magnitude and direction of the kick produced in the
  second SN (Appendix~\ref{A:kick}),
\item[iv)] the angles $\varphi^\prime$ and $\varpi$ specifying the
  directions of the spins $\mathbf{S^{\prime}}$ and
  $\mathbf{S^{\prime\prime}}$ before the second SN
  (Section~\ref{SS:example}),
\item[v)] the angle $\Delta \Phi$ between the projections of the BH
  spins in the orbital plane at separation $a_{\rm PNi}$.

\end{itemize}
The angles $\varphi^\prime$, $\varpi$, and $\Delta
\Phi$ in items iv) and v) above are uniformly distributed in the range
    [$0, 2\pi$].  The synthetic populations generated in this
    procedure determine the initial conditions for the PN equations of
    motion described in Section~\ref{PNinspiral}.

A binary-star system can {\it fail} to produce a merging BH binary for
one of the following reasons:

\begin{itemize}

\item[i)] it is unbound by the first SN ($e_1 > 1$);
\item[ii)] it merges during the CE evolution between the two SN
  ($a_{1CE} < a_{\rm mCE}$);
\item[iii)] it is unbound by the second SN ($e_2 > 1$);
\item[iv)] the time $t$ required for gravitational radiation to shrink
  the semimajor axis from $a_2$ to $a_{\rm PNi}$, found by solving the
  coupled PN equations (\ref{petersdtda}) and (\ref{petersdeda}),
  exceeds the Hubble time $t_{\rm H} \simeq 10^{10}$~Gyr.

\end{itemize}
Table~\ref{rates} lists the fraction of simulated binaries $\nu_{\rm
  SN1}$, $\nu_{\rm mCE}$, $\nu_{\rm SN2}$, and $\nu_{H}$ that fail to
produce merging BH binaries for reasons i) through iv) listed above,
as well as the fraction $\nu_{\rm BH} =1-(\nu_{\rm SN1}+\nu_{\rm mCE}
+ \nu_{\rm SN2} + \nu_{\rm H})$ that {\it do} evolve into such
binaries.

\begin{figure*}
\begin{tabular*}{\textwidth}{c@{\extracolsep{\fill}}c}
\includegraphics[width=0.48\textwidth]{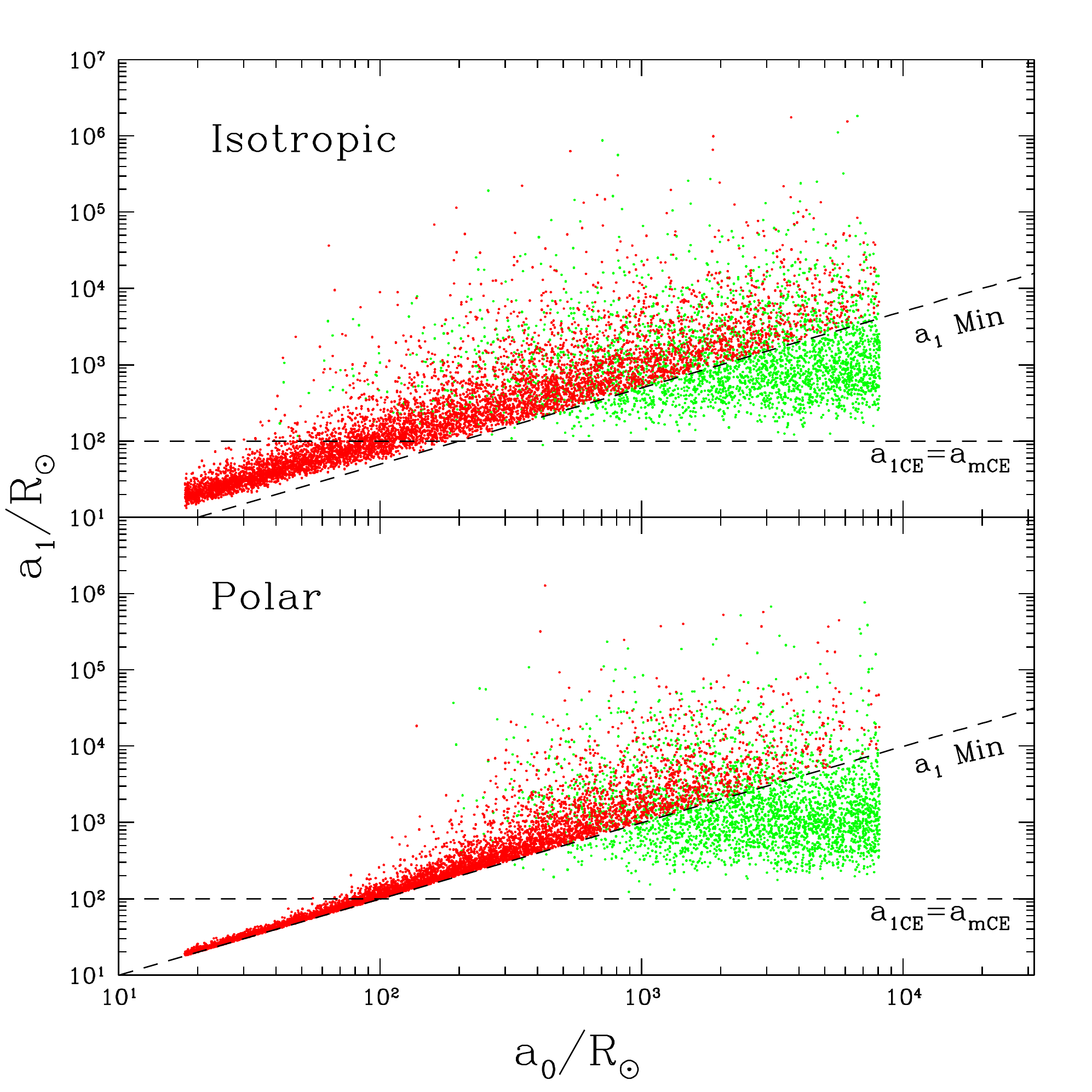}
&
\includegraphics[width=0.48\textwidth]{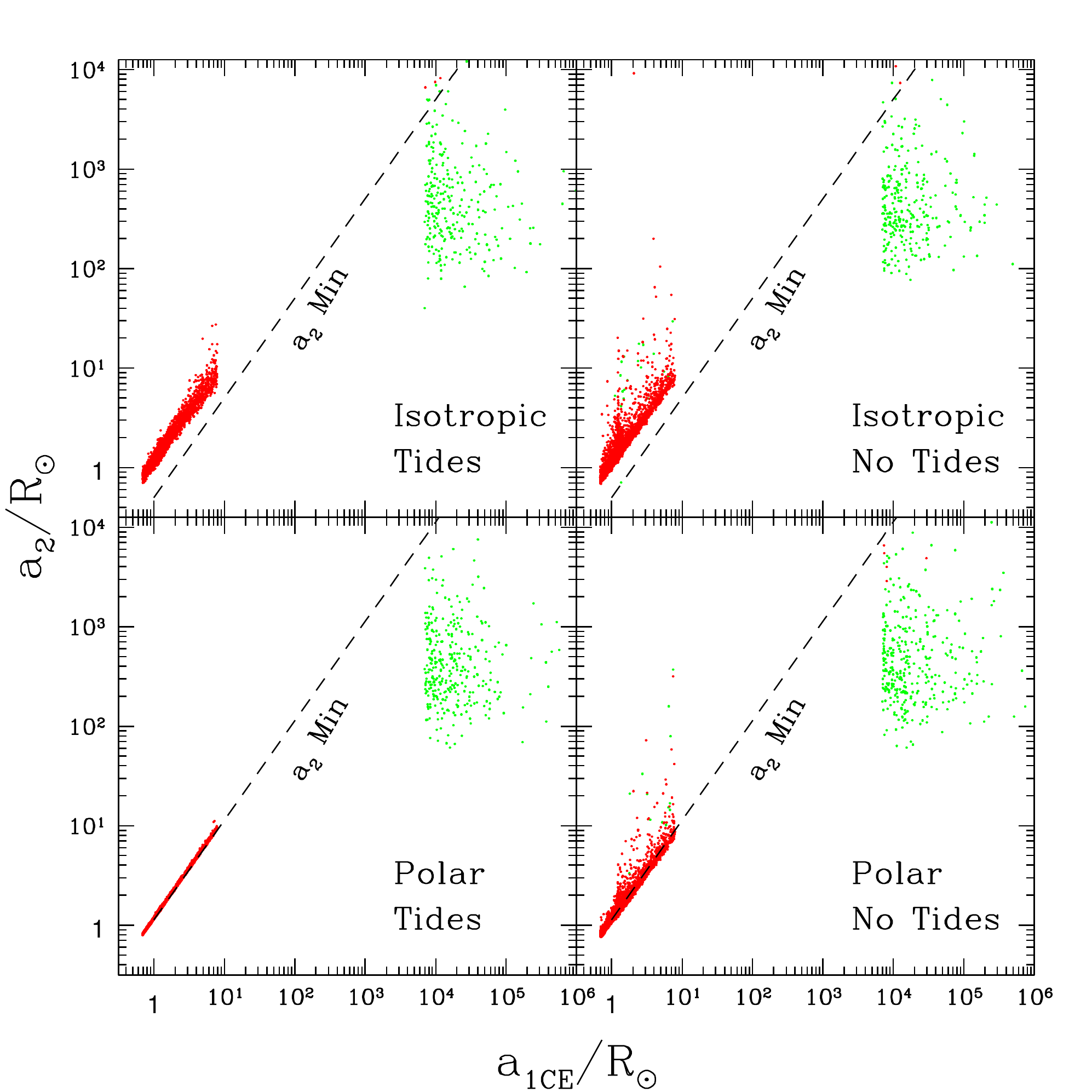}
\end{tabular*}
\caption{(Color online.) Scatter plot showing the change in the
  semimajor axis due to the first (left panel: $a_0 \to a_1$) and
  second (right panel: $a_{\rm 1CE} \to a_2$) SN. All plots refer to
  the SMR scenario, but the behavior in the RMR scenario is very
  similar. Darker (red) dots represent binaries that remain bound
  after each explosion, while lighter (green) dots correspond to
  binaries that are unbound. Dashed lines show the minimum post-SN
  semimajor axis $a_{f, {\rm Min}}$ given by Eq.~(\ref{E:afmin}) and
  the critical semimajor axis $a_{\rm mCE}$ given by Eq.~(\ref{E:mCE})
  below which binaries merge during CE evolution.  Kicks are too small
  to saturate the isotropic limit $a_{f, {\rm Min}}$ for $a_i \lesssim
  10^2 R_\odot$.}
\label{kicksep}
\end{figure*}

The failure fractions indicate the relative importance of different
physical phenomena.  To emphasize the sensitivity of our results to
the highly uncertain SN kicks, we also show how these fractions change
when the BH kick $v_{\rm BH} = (1-f_{\rm fb})v_{\rm pNS}$ fully equals
that imparted to the protoneutron star ($f_{\rm fb} = 0$) rather than our
canonical choice ($f_{\rm fb} = 0.8$); see Appendix~\ref{A:kick} for more
details.  Stronger kicks unbind more binaries during the first SN,
increasing $\nu_{\rm SN1}$ and thereby reducing the overall fraction
$\nu_{\rm BH}$ of binaries that survive to form BH binaries.  This
qualitatively agrees with results of detailed population-synthesis
models; see models S, V8, and V9 in \cite{2012arXiv1208.0358B}.  We
adopt $f_{\rm fb} = 0.8$ in the remainder of the paper.

Fig.~\ref{kicksep} shows how the choices that define our fiducial
scenarios affect whether SN kicks unbind the binaries.  One result
apparent from this plot (and supported by the failure fractions
$\nu_{\rm SN}$ listed in Table~\ref{rates}) is that the probability of
unbinding the system depends only weakly on whether the SN kicks are
isotropic or polar.  This is consistent with the findings of
\cite{2008MNRAS.384.1393P}, which suggest mild sensitivity to
$\theta_b$ when the typical kick velocity $v_{\rm BH} \sim 50$~km/s is
small compared to the orbital velocity $v_0 \simeq 2.4 \times 10^3
(M/30M_\odot)^{1/2}(a/R_\odot)^{-1/2}$~km/s.  Fig.~\ref{kicksep} also
shows the effect of tides on the fraction $\nu_{\rm BH}$ of BH
binaries produced.  In the absence of tidal dissipation (``No
Tides''), the binaries have nonzero eccentricity ($e_i \neq 0$) when
the second SN occurs.  Eq.~(\ref{Ksemimajoraxis}) shows that the final
semimajor axis $a_f$ has additional dependence on the true anomaly
$\psi_i$ in this limit, broadening the distribution of $a_f$, as can
be seen in the right panel of Fig.~\ref{kicksep}.  The kicks can add
coherently to the large orbital velocities near pericenter of highly
eccentric orbits, allowing binaries to become unbound even after CE
evolution has reduced the semimajor axis: cf. the handful of
light-gray (green) points with $a_{1CE} \lesssim 10 R_\odot$ in the
right panel of Fig.~\ref{kicksep}).  This increases the fraction
$\nu_{\rm SN2}$ of binaries unbound in the second SN when tides are
``Off'' in Table~\ref{rates}.  The importance of CE evolution can be
seen as well: virtually all binaries that fail to form a CE ($a_{1CE}
\gtrsim 10^4 R_\odot$) are unbound by the second SN.  Binaries bound
tightly enough to survive the second SN almost always manage to merge
through GW emission in less than a Hubble time ($\nu_{\rm H} \ll 1$).

\section{Post-Newtonian Inspiral} 
\label{PNinspiral}

\subsection{Post-Newtonian equations of motion} 

At large orbital separations, the dynamics of BH binaries in vacuum
can be approximated by expanding the Einstein equations in a
perturbative PN series, where the perturbative parameter is the ratio
$v/c$ of the orbital velocity to the speed of light.  For historical
reasons, one usually says that a quantity is expanded up to $k$PN
order if all terms up to order $(v/c)^{2k}$ are retained.  Following
common practice in the general relativity literature, in this Section
we will use geometrical units such that $G=c=1$.

The PN approximation can describe the evolution of stellar-mass
binaries down to separations $a\sim 10 M$ (i.e. $a\sim 10^{-4} R_\sun$
for a BH binary with $M=10M_\sun$), beyond which fully nonlinear
numerical simulations are needed
\cite{2002LRR.....5....3B,Yunes:2008tw,2009PhRvD..80h4043B,Zhang:2011vha}.
GW detection templates depend on the binary parameters when the system
enters the sensitivity band of the detectors, which is well into the
regime where PN corrections are significant, but astrophysical models
of BH evolution (as implemented e.g. in population-synthesis codes)
have so far neglected all general-relativistic effects.  The main goal
of this Section is to show that solving the PN equations of motion is
necessary to determine the orientation of BH spins when binaries enter
the sensitivity band of GW detectors such as Advanced LIGO/Virgo and
the Einstein Telescope.

The PN equations of motion and gravitational waveforms for spinning BH
binaries were derived by several authors (see
e.g.~\cite{1995PhRvD..52..821K,PhysRevD.79.104023,PhysRevD.84.049901}).
Our previous investigations of spin dynamics considered binaries on
circular orbits; as shown in Section \ref{PNinit} below, this is an
excellent approximation for most binaries in our sample.  They also
included high-order PN terms such as the monopole-quadrupole
interaction and the spin-spin self interactions
\cite{2010PhRvD..81h4054K,2010ApJ...715.1006K,2012PhRvD..85l4049B},
that we report for completeness below.

For circular orbits with radius $a$ and orbital velocity
$v=(GM/a)^{1/2}$, the ``intrinsic'' dynamics of a binary system
depends on 10 variables: the two masses $(m_1,\,m_2)$, the spins
$\mathbf S_1$ and $\mathbf S_2$ and the direction of the orbital
angular momentum $\hat {\mathbf{L}}$. At the PN order we consider both
spin magnitudes and the mass ratio $q$ remain fixed during the
inspiral.  This leaves 7 independent degrees of freedom. Because BHs
are vacuum solutions of the Einstein equations, there is only one
physical scale in the problem (the total mass of the binary
$M$). Rescaling all quantities relative to the mass $M$, we are left
with 6 ``intrinsic'' parameters.

It is convenient to analyze the precessional dynamics in the frame
where the direction of the orbital momentum $\mathbf{\hat L}$ lies
along the $z$-axis. If we take (say) the $x$-axis to be oriented along
the projection of $\mathbf{S_1}$ on the orbital plane (see Fig.~1 in
\cite{2004PhRvD..70l4020S}), we are effectively imposing 3 additional
constraints just by our choice of the reference frame (2 components of
$\mathbf{\hat L}$ and 1 component of $\mathbf{S_1}$ are set equal to
zero). Then the only 3 variables describing precessional dynamics are
the angles $\theta_1$, $\theta_2$ and $\Delta\Phi$, as defined in
Eqs.~(\ref{theta1theta2}) and (\ref{dphitheta12}). The angle between
the two spins $\theta_{12}$ is related to the other independent
variables as follows:
\begin{align}
\cos\theta_{12}=\sin\theta_1 \sin\theta_2 \cos\Delta\Phi +\cos\theta_1 \cos\theta_2.
\label{deftheta12}
\end{align}

In summary, for any given binary with intrinsic parameters
($q$,\,$\chi_1$,\,$\chi_2$), the precessional dynamics is encoded in
the variables ($\theta_1$,\,$\theta_2$,\,$\Delta\Phi$) as functions of
the orbital velocity $v$ or (equivalently) of the orbital frequency
$\omega = v^3/M$.  These variables can be evolved forward in time by
integrating the following PN equations of motion:
\begin{widetext}
\begin{alignat}{2}
\label{precession1}
\frac{d \mathbf{ S_1}}{dt} &= \mathbf{\Omega_1}\times \mathbf{ S_1}, 
\qquad \qquad
M \mathbf{\Omega_1}&= \eta v^5 \left( 2 +\frac{3q}{2} \right) \mathbf{\hat L}
+  \frac{v^6}{2M^2}\left[  \mathbf{ S_2} - 3 \left(\mathbf{\hat L} \cdot \mathbf{ S_2} \right) \mathbf{\hat L}
- 3q \left(\mathbf{\hat L} \cdot \mathbf{ S_1} \right) \mathbf{\hat L}\right];
\\ \label{precession2}
\frac{d \mathbf{ S_2}}{dt} &= \mathbf{\Omega_2}\times \mathbf{ S_2}, 
\qquad \qquad
M \mathbf{\Omega_2}&= \eta v^5 \left( 2 +\frac{3}{2q} \right) \mathbf{\hat L}
+  \frac{v^6}{2M^2}\left[\mathbf{ S_1} - 3 \left(\mathbf{\hat L} \cdot \mathbf{ S_1} \right) \mathbf{\hat L}
- \frac{3}{q} \left(\mathbf{\hat L} \cdot \mathbf{ S_2} \right) \mathbf{\hat L}\right];
\end{alignat}
\begin{align}
\label{angmom}
\frac{d \mathbf{\hat L}}{dt} = -\frac{v}{\eta M^2}
\frac{d }{dt} ( \mathbf{ S_1}+ \mathbf{ S_2});&
\end{align}
\begin{align}
\frac{dv}{dt}&= \frac{32}{5}\frac{\eta}{M}v^9 \Bigg\{ 1-
 v^2 \frac{743 +924 \eta }{336}+ 
 v^3 \Bigg[ 4 \pi - \sum_{i=1,2}
  \chi_i (\mathbf{\hat S_i} \cdot\mathbf{\hat L}   )\left(\frac{113}{12}\frac{m_i^2}{M^2} + \frac{25}{4}\eta\right)  \Bigg]
 \notag \\
 &+  v^4
\Bigg[\frac{34103}{18144}+\frac{13661}{2016}\eta +\frac{59}{18}\eta^2  
+ \frac{\eta \chi_1 \chi_2}{48} \left( 721(\mathbf{\hat S_1} \cdot\mathbf{\hat L})(\mathbf{\hat S_2} \cdot\mathbf{\hat L})
-247 (\mathbf{\hat S_1}\cdot\mathbf {\hat S_2} ) \right)
\notag \\ 
& + \frac{1}{96} \sum_{i=1,2}
\left(\frac{m_i \chi_i}{M}\right)^2 \left( 719 (\mathbf{\hat S_i} \cdot\mathbf{\hat L})^2 - 233 \right)
\Bigg]-
 v^5 \pi \frac{4159 +15876 \eta}{672} 
  \notag \\
&+ v^6\Bigg[\frac{16447322263}{139708800}+\frac{16}{3}\pi^2 -\frac{1712}{105}\left(\gamma_E +\ln4v\right)+ 
\left( \frac{451}{48} \pi^2 - \frac{56198689}{217728} \right) \eta
+\frac{541}{896}\eta^2 -\frac{5605}{2592}\eta^3\Bigg]
\notag \\
&+ v^7 \pi \Bigg[ -\frac{4415}{4032}+\frac{358675}{6048}\eta +\frac{91495}{1512}\eta^2\Bigg]
+O(v^8)\Bigg\} ; 
\label{radiationreaction}
\end{align}
\end{widetext}
where $\eta= m_1 m_2/ M^2$ and $\gamma_E\simeq0.577$ is Euler's
constant.

The leading terms in Eqs.~(\ref{precession1})-(\ref{precession2}), up
to ${\cal O}(v^5)$ or 2.5PN order, describe precessional motion about
the direction of the orbital angular momentum $\mathbf{\hat L}$.  We
assumed that these terms dominated during the PN inspiral of the
previous Section, allowing $\gamma'_2$ and $\gamma_2^{\prime\prime}$
to remain fixed at $a > 1000M$.  Spin-orbit couplings appear at 3PN,
and they are the reason for the existence of the resonant
configurations \cite{2004PhRvD..70l4020S}.  From Eq.~(\ref{angmom}) we
see that the direction of the angular momentum evolves on a
precessional timescale, while Eq.~(\ref{radiationreaction}) implies
that its magnitude decreases on the (longer) radiation-reaction
timescale due to GW emission. The leading (quadrupolar) order of
Eq.~(\ref{radiationreaction}) is equivalent to the circular limit of
Eq.~(\ref{petersdtda}) when we recall that $v^2=M/a$.
 
Higher-order PN terms in the equations of motion were recently
computed \cite{favata}. We modified
Eqs.~(\ref{precession1})-(\ref{radiationreaction}) to include these
new terms, finding that they affect the late-time dynamics of
individual binaries but have negligible influence on the statistical
behavior of our samples. The robustness of these statistical
properties under the inclusion of higher-order PN terms was already
noted in
\cite{2012PhRvD..85l4049B,2010PhRvD..81h4054K,2010ApJ...715.1006K}.
For completeness we retained the higher-order PN terms that will be
reported in \cite{favata} in our Monte Carlo simulations, but we
stress again that they have no observable impact on our results.

At a given separation $a$, Schnittman's resonant configurations can be
found by forcing the three vectors ${\mathbf S_1}$, ${\mathbf S_2}$
and $\hat{\mathbf{L}}$ to lie in a plane
($\Delta\Phi=0^\circ,\pm180^\circ$) and by imposing the constraint
that the second time derivative of $\cos \theta_{12}$ vanish
\cite{2004PhRvD..70l4020S}.  A one-parameter family of configurations
with $\Delta\Phi=0^\circ$ and $\theta_1<\theta_2$ satisfies this
resonant constraint, as does a second one-parameter family with
$\Delta\Phi=\pm180^\circ$ and $\theta_1>\theta_2$.  As $a$ decreases
due to GW emission, the curves determined by these one-parameter
families change, sweeping through a large region of the
$(\theta_1,\,\theta_2)$ parameter space. The resonant constraint
evolves toward the diagonal $\theta_1=\theta_2$ as $a \to 0$.
Individual resonant binaries move towards the diagonal in the
$(\theta_1,\,\theta_2)$ plane along trajectories over which the
projection $\mathbf{S_0} \cdot \mathbf{\hat L}$ of the spin
combination $\mathbf{S_0}$ defined in the effective-one-body model
\cite{2001PhRvD..64l4013D},
\begin{align}
\mathbf{S_0} = (1+q) \mathbf{S_1} + (1+q^{-1}) \mathbf{S_2},
\label{S0}
\end{align}
is approximately constant (cf.~Figs.~1 and 2 of
\cite{2010PhRvD..81h4054K}).
Resonant configurations with $\Delta\Phi=0$ tend to align the two
spins with each other, so that $\theta_{12} \to 0^\circ$ near
merger. On the other hand, configurations with
$\Delta\Phi=\pm180^\circ$ identified by their constant value of
$\mathbf{S_0} \cdot \mathbf{\hat L}$ evolve towards
\begin{align}
\cos\theta_{12} \to 2\left[\frac{ (1+q) \mathbf{S_0}\cdot\mathbf{\hat L}} {(\chi_1 + q\chi_2)M^2}\right]^2 -1~.
\label{constrainth12}
\end{align}

\subsection{Initial conditions for the PN evolution} 
\label{PNinit}

By construction, all of the merging BH binaries produced in
Section~\ref{binformation} have $M=13.5 M_\sun$, $q=0.8$, and
$\chi_1=\chi_2=1$.  For this mass ratio and these spin magnitudes,
binaries become attracted towards resonances (``resonant locking'') at
separations $a \lesssim 100M$ \cite{2004PhRvD..70l4020S}.  Previous
studies suggest that the spin-orbit resonances remain influential
provided $q \gtrsim 0.4$ and $\chi_i \gtrsim 0.5$
\cite{2010PhRvD..81h4054K,2010ApJ...715.1006K,2012PhRvD..85l4049B}.
To be safe, we begin following binaries at an initial separation
$a_{\rm PNi}=1000M$ large enough so that we can neglect spin-spin
coupling at greater separations \cite{2010PhRvD..81h4054K}. Recall
that the mass ratio was defined such that $q \equiv m_2/m_1 \leq 1$.
In the SMR scenario, the primary yields the larger BH ($M'_{\rm
  BH}>M''_{\rm BH}$), so the angles are initialized to be
\begin{equation} \label{SMRtheta}
\theta_1 = \gamma'_2, \qquad \theta_2 = \gamma''_2~.
\end{equation}
In the RMR case, the primary transfers so much mass to the secondary
prior to the first SN that it actually produces the smaller BH
($M'_{\rm BH}<M''_{\rm BH}$), implying that we must reverse our
initialization:
\begin{equation} \label{RMRtheta}
\theta_1 = \gamma''_2, \qquad \theta_2 = \gamma'_2~.
\end{equation}
Although our decision to neglect spin-spin coupling for $a > a_{\rm
  PNi}$ allows us to initialize $\theta_i$ in this manner, the
lower-order spin-orbit coupling allows $\Delta \Phi$ to evolve on the
precessional timescale, which is short compared to the time it takes
to inspiral from $a_2$ to $a_{\rm PNi}$.  We can therefore choose
$\Delta \Phi$ at $a_{\rm PNi}$ to be uniformly distributed in the
range $[-180^\circ, +180^\circ]$.  Finally, since gravitational
radiation is very efficient at circularizing the orbit [to leading
  order $e \propto a^{19/12}$; see Eq.~(\ref{petersdeda})], we assume
that all BH binaries have circularized by the time they reach $a_{\rm
  PNi}$.  We checked this assumption by numerically integrating
Eq.~(\ref{petersdeda}) from $a_2$ to $a_{\rm PNi}$ after initializing
it with the values $e_2$ predicted following the second SN; the
residual eccentricity at $a_{\rm PNi}$ was less than $10^{-4}$ for all
BH binaries in our samples.

\begin{figure*}[htb]
\begin{tabular*}{\textwidth}{c@{\extracolsep{\fill}}c}
\includegraphics[width=\figw]{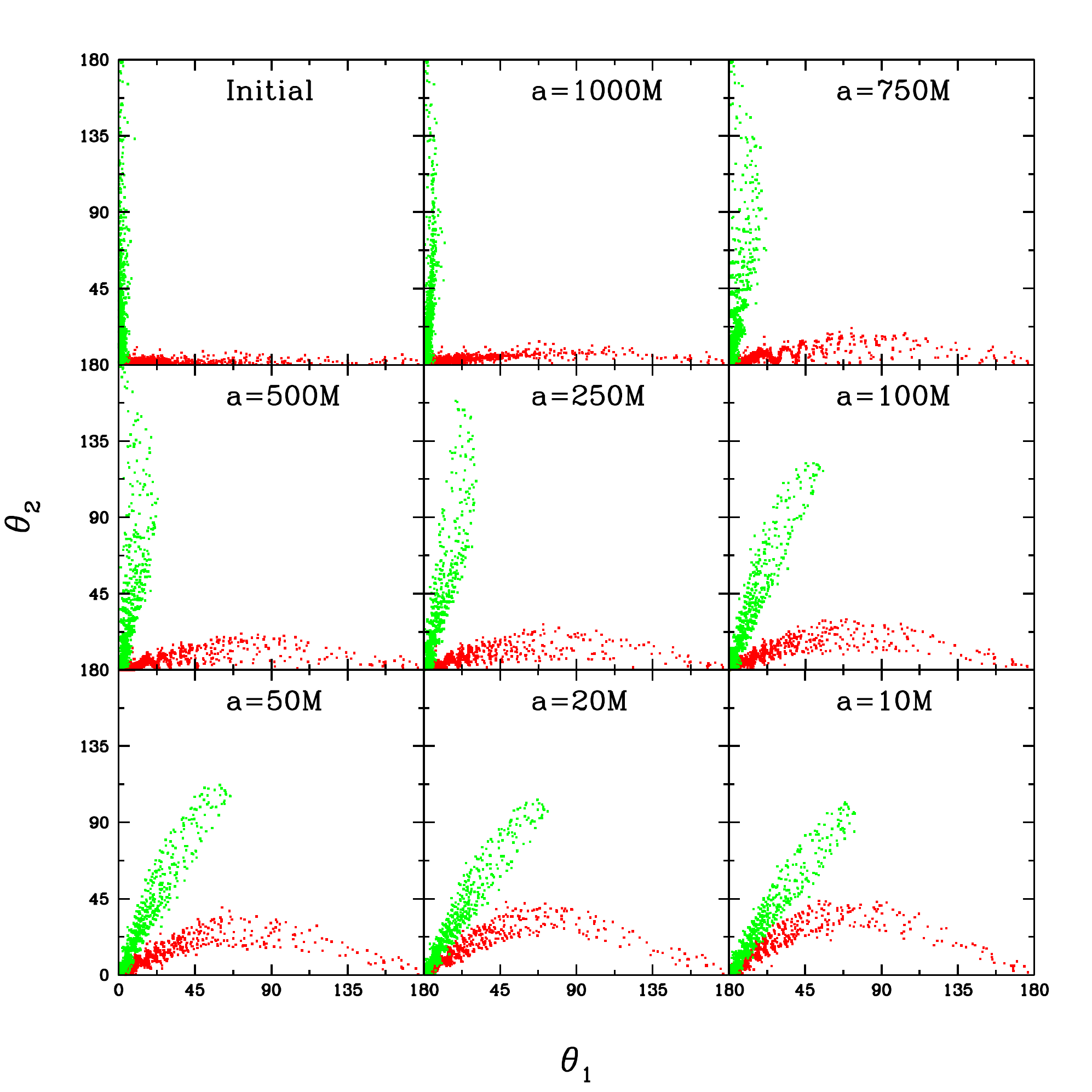} 
&
\includegraphics[width=\figw]{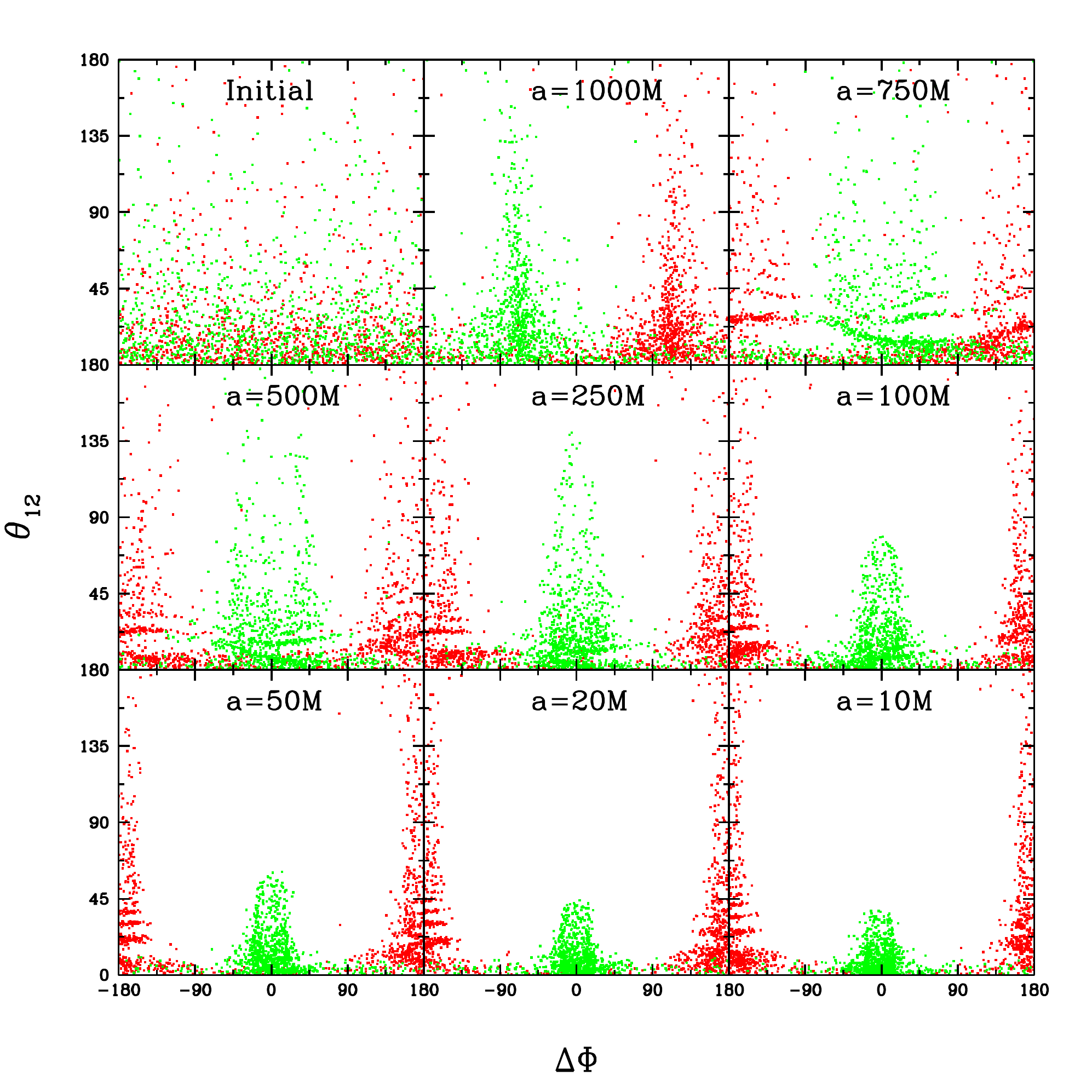}
\end{tabular*}
\caption{(Color online.) Scatter plots of the PN inspiral of maximally
  spinning BH binaries with mass ratio $q=0.8$ from an initial
  separation $a_{\rm PNi}$ just above $1000M$ to a final separation
  $a_{\rm PNf} = 10M$.  The left panel shows this evolution in the
  $(\theta_1,\theta_2)$ plane and the right panel shows the evolution
  in the $(\Delta\Phi,\theta_{12})$ plane. Darker (red) and lighter
  (green) dots refer to the SMR and RMR scenarios, respectively. The
  initial distribution for these Monte Carlo simulations was
  constructed from an astrophysical model with efficient tides and
  isotropic kicks. An animated version of this plot is available
  online at the URL:
  \\\url{http://www.phy.olemiss.edu/~berti/tides_isotr.gif}}
\label{PNtidesiso}
\end{figure*}

\begin{figure*}[htb]
\begin{tabular*}{\textwidth}{c@{\extracolsep{\fill}}c}
\includegraphics[width=\figw]{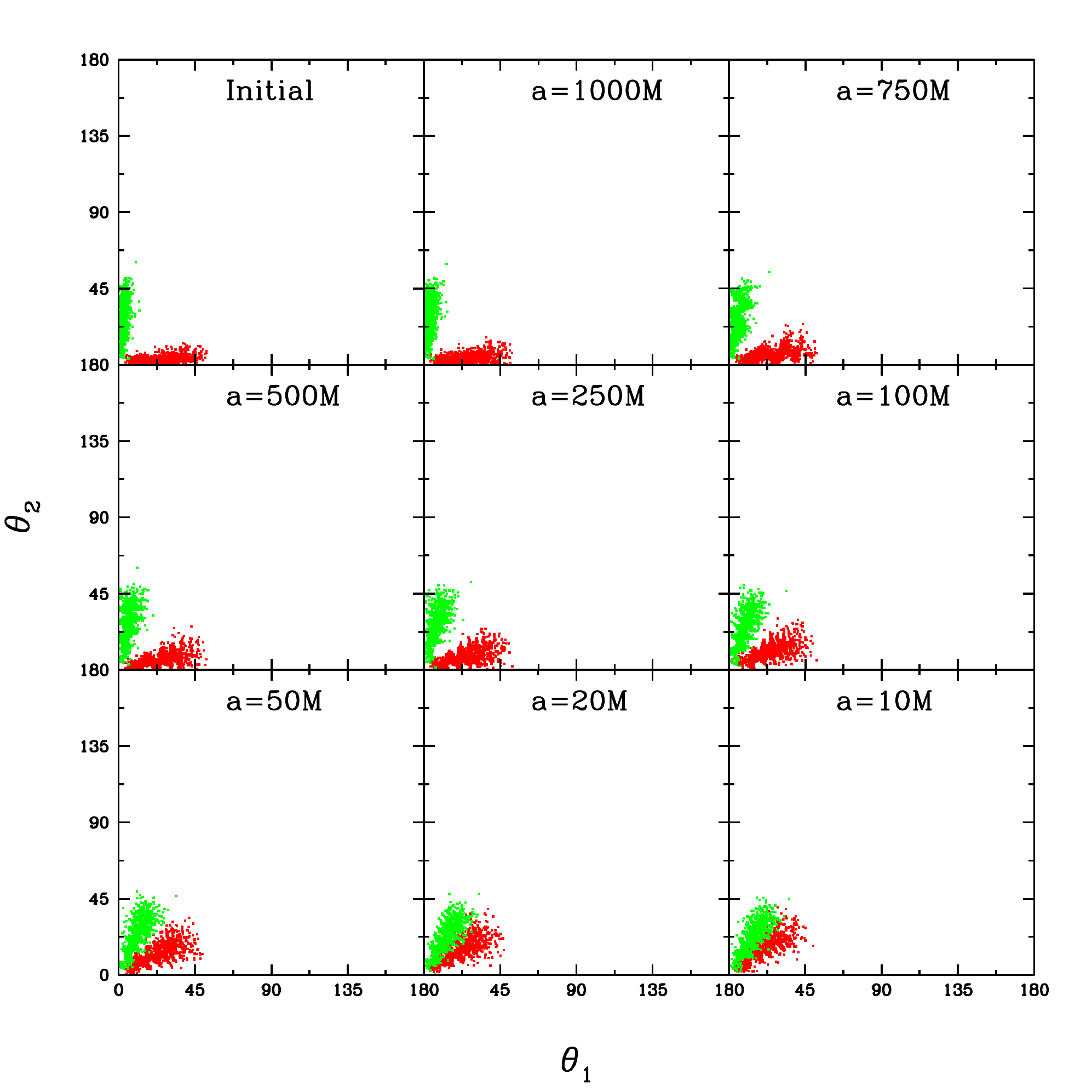} 
&
\includegraphics[width=\figw]{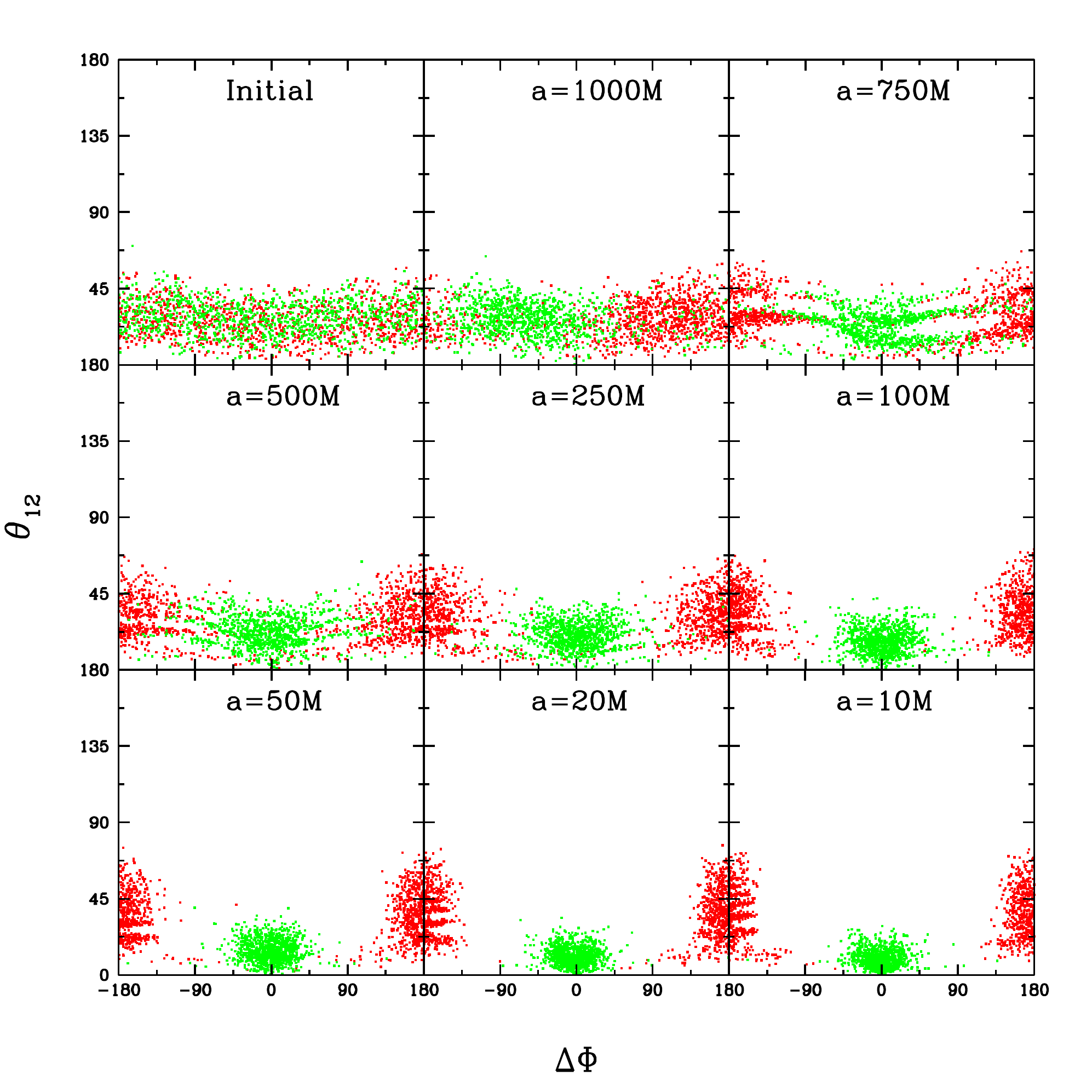}
\end{tabular*}
\caption{(Color online.) Scatter plots of the same quantities shown in
  Fig.~\ref{PNtidesiso} for an astrophysical model with efficient
  tides and polar kicks. For an animated version of this plot, see:
  \url{http://www.phy.olemiss.edu/~berti/tides_polar.gif}}
\label{PNtidespolar}
\end{figure*}

\begin{figure*}[htb]
\begin{tabular*}{\textwidth}{c@{\extracolsep{\fill}}c}
\includegraphics[width=\figw]{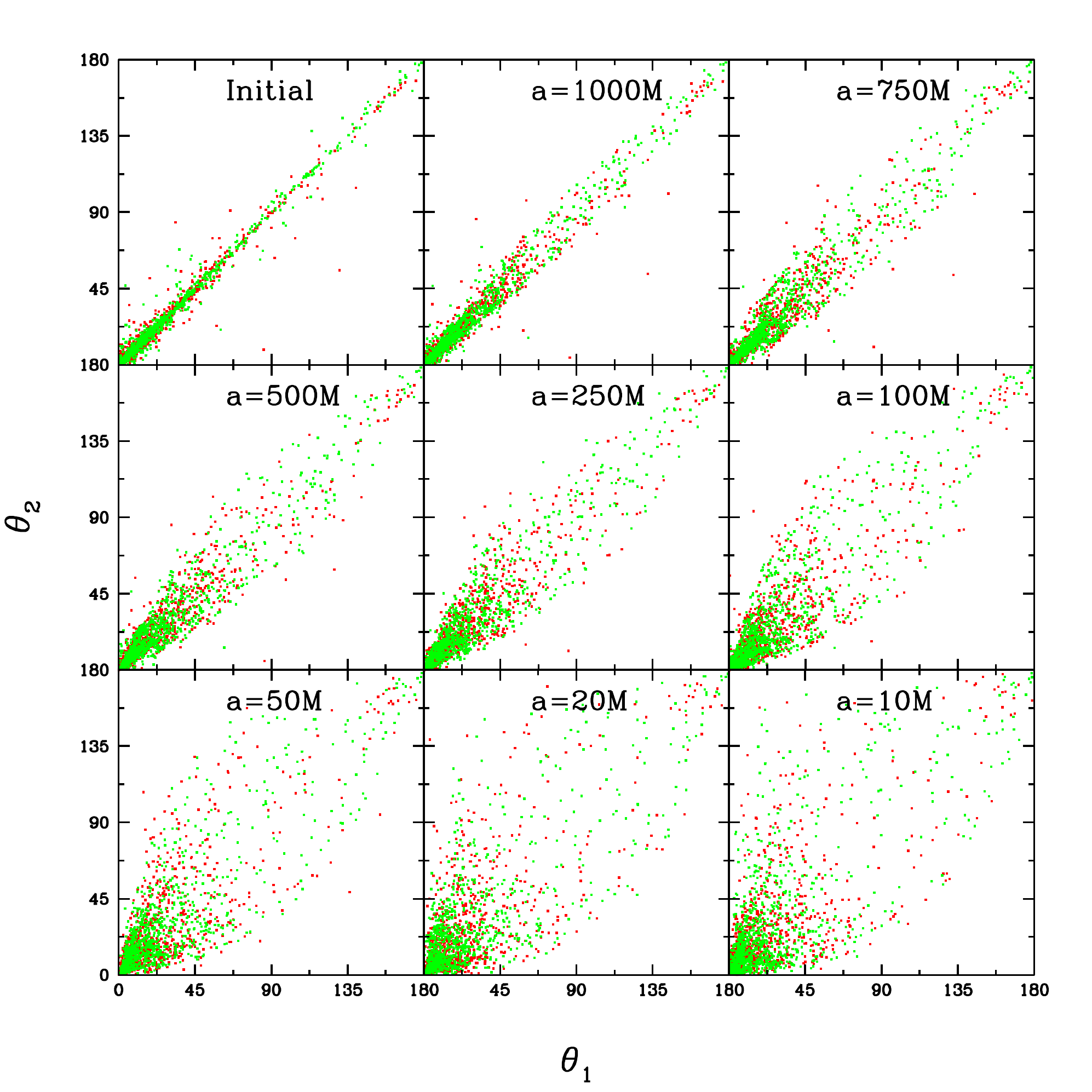} &
\includegraphics[width=\figw]{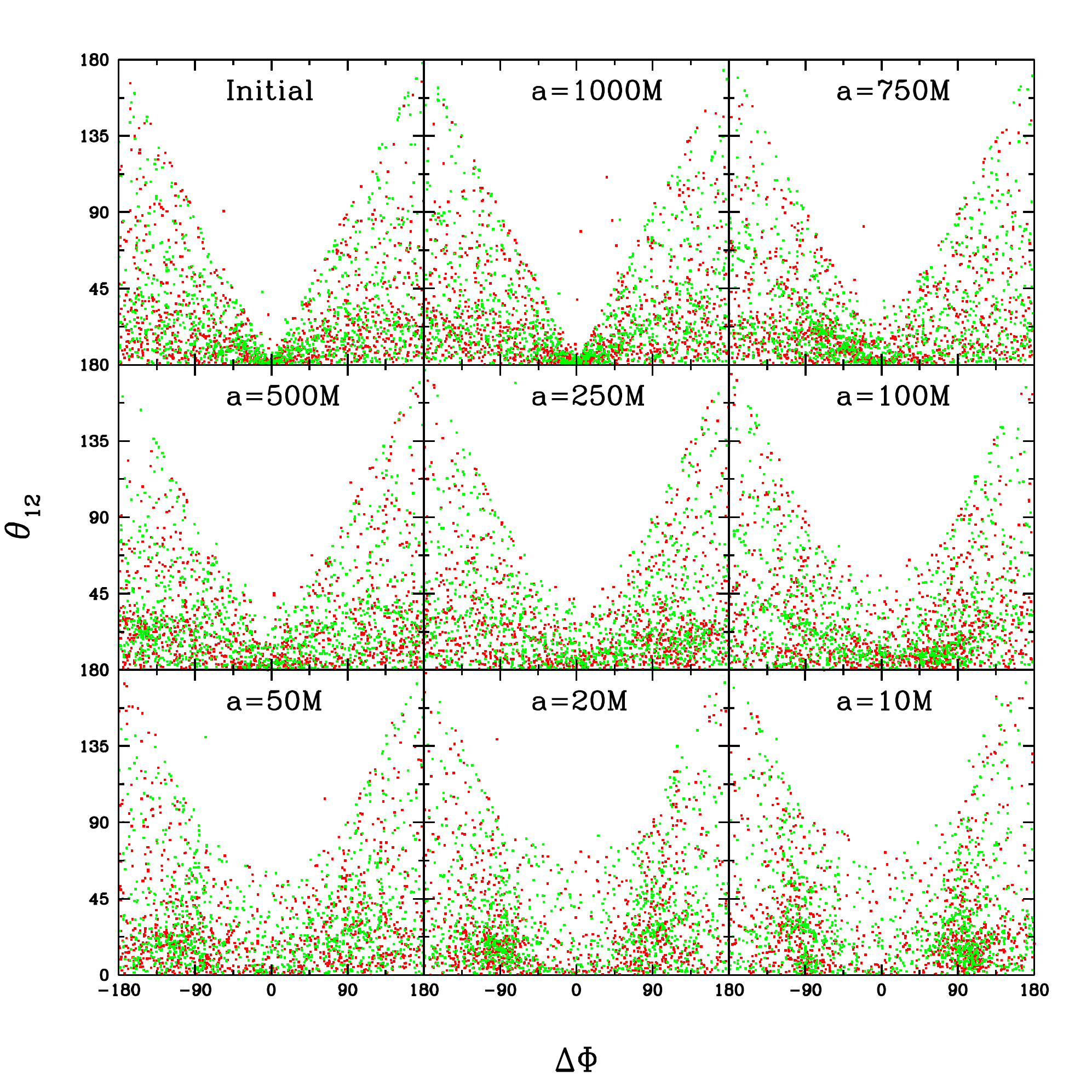} 
\end{tabular*}
\caption{Scatter plots of the same quantities shown in
  Fig.~\ref{PNtidesiso} for an astrophysical model with inefficient
  tides and isotropic kicks. For an animated version of this plot,
  see: \url{http://www.phy.olemiss.edu/~berti/notides_isotr.gif}}
\label{PNnotidesiso}
\end{figure*}

\begin{figure*}[htb]
\begin{tabular*}{\textwidth}{c@{\extracolsep{\fill}}c}
\includegraphics[width=\figw]{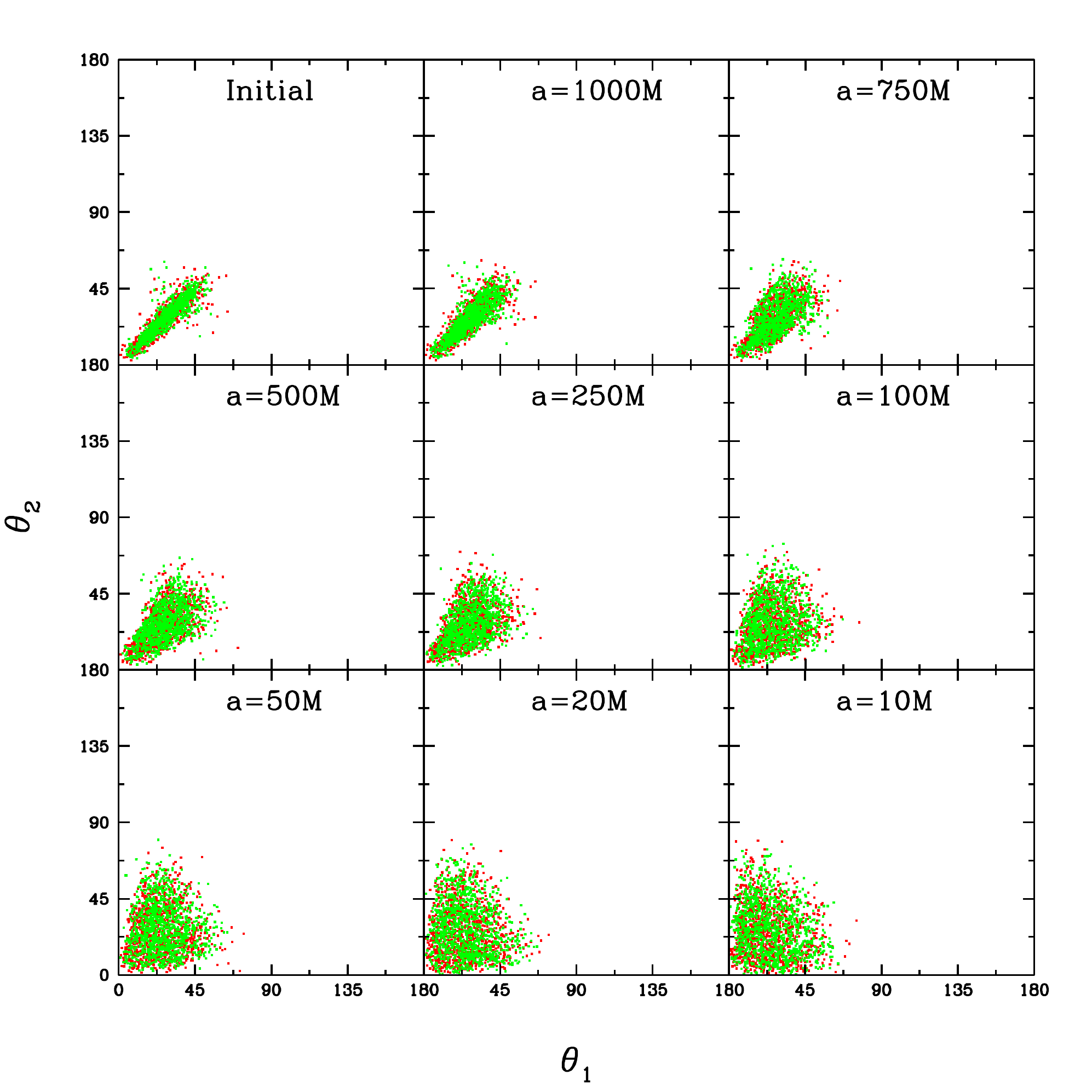} &
\includegraphics[width=\figw]{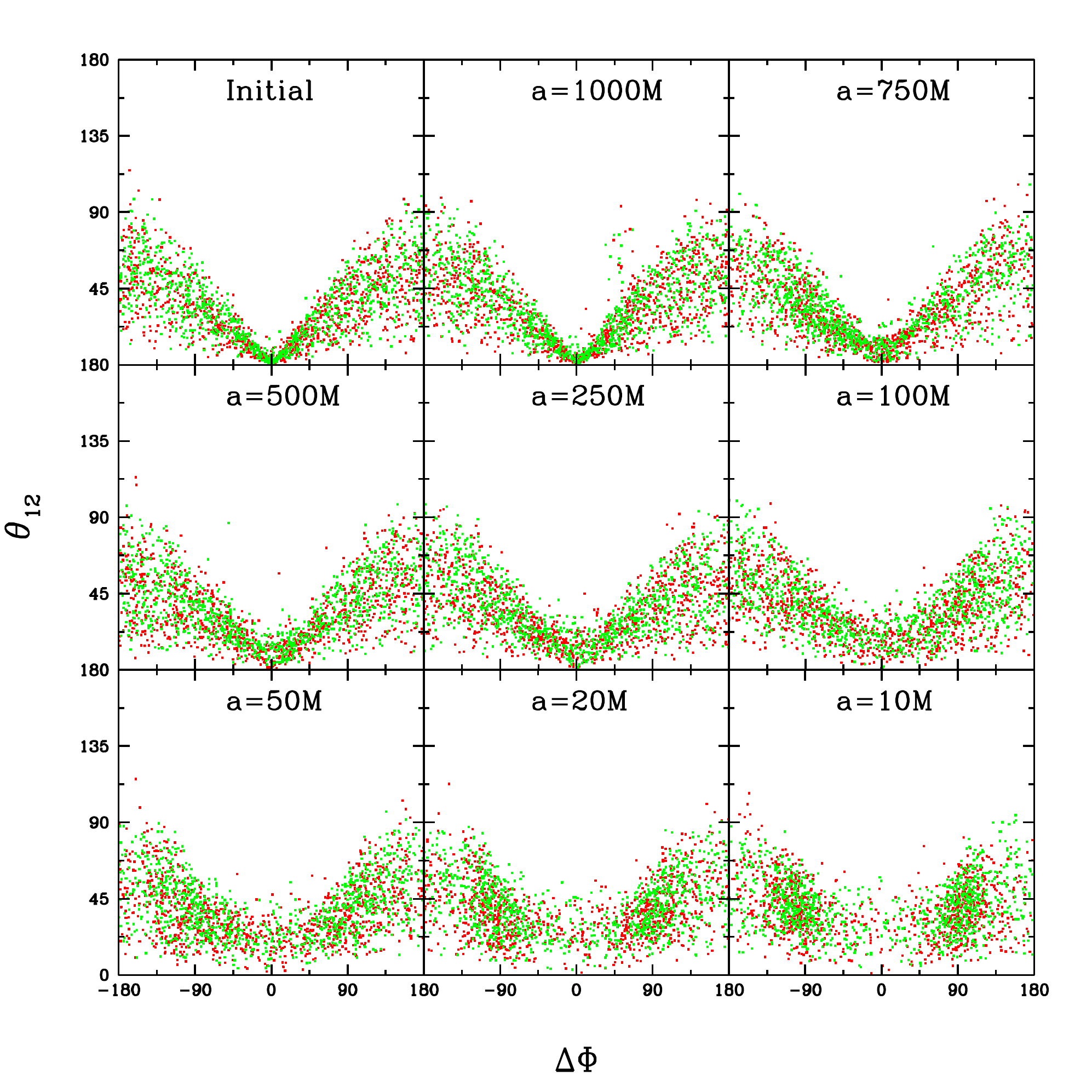} 
\end{tabular*}
\caption{Scatter plots of the same quantities shown in
  Fig.~\ref{PNtidesiso} for an astrophysical model with inefficient
  tides and polar kicks. For an animated version of this plot, see:
  \url{http://www.phy.olemiss.edu/~berti/notides_polar.gif}}
\label{PNnotidespolar}
\end{figure*}

\subsection{Results}

We evolved $10^3$ BH binaries for each of the 8 different fiducial
astrophysical scenarios described in
Section~\ref{sec:sub:RandomEvolve} from an initial
separation\footnote{The $a=1000M$ snapshots in the figures of this
  Section are taken shortly after the beginning of the PN
  evolution. The angle $\Delta\Phi$ varies on the precessional
  timescale and can therefore change quite rapidly before the
  separation decreases appreciably on the longer inspiral timescale.
  The initial clustering in $\Delta\Phi$ visible in the top-right
  panels of Figs.~\ref{PNtidesiso} and \ref{PNtidespolar} is {\em not}
  a resonant effect, as the binaries continue to sweep through all
  values of $\Delta\Phi$ at these large separations.  It results
  instead from the different rates at which binaries in the SMR and
  RMR populations precess, segregating the groups from each other
  during the first few precessional cycles. This behavior is better
  illustrated by the animations available online at the following
  URLs, which refer to efficient tides with isotropic kicks, efficient
  tides with polar kicks, inefficient tides with isotropic kicks, and
  inefficient tides with polar kicks, respectively:
  \\\url{http://www.phy.olemiss.edu/~berti/tides_isotr.gif}
  \\\url{http://www.phy.olemiss.edu/~berti/tides_polar.gif}
  \\\url{http://www.phy.olemiss.edu/~berti/notides_isotr.gif}
  \\\url{http://www.phy.olemiss.edu/~berti/notides_polar.gif}} $a_{\rm
  PNi} = 1000M$ to a final separation $a_{\rm PNf}=10M$.  This final
separation roughly indicates where the PN approximation breaks down
and full numerical relativity becomes necessary
\cite{2002LRR.....5....3B,Yunes:2008tw,2009PhRvD..80h4043B,Zhang:2011vha}.
To reduce the Poisson noise in the histograms of Fig.~\ref{intro_phi},
we used larger samples of $10^4$ BH binaries.  We integrated the PN
equations (\ref{precession1})-(\ref{radiationreaction}) using a
\textsc{stepperdopr5} integrator in \textsc{C++}
\cite{2002nrc..book.....P}, progressively refining the time steps at
small separations (see \cite{2010PhRvD..81h4054K} for further
details).

In Fig.~\ref{PNtidesiso}, we show the evolution of the dynamical
variables ($\theta_1, \theta_2, \Delta \Phi$) for both the SMR and RMR
scenarios with efficient tides and isotropic kicks.  As already
anticipated in the introduction, efficient tidal interactions lead to
spin orientations that are strongly affected by spin-orbit resonances.
When binaries are brought close enough to resonant configurations by
precessional motion and gravitational-radiation reaction, they no
longer precess freely through all values of $\Delta \Phi$, but instead
oscillate about the resonant configurations
\cite{2004PhRvD..70l4020S,2010PhRvD..81h4054K}.  In the SMR scenario,
the initial orientation of the spins is such that $\theta_1>\theta_2$,
and the binaries lock into resonances with $\Delta\Phi = \pm180^\circ$
[darker (red) points in Fig.~\ref{PNtidesiso}].  In contrast, in the
RMR scenario the initial spins have $\theta_1<\theta_2$ and the
binaries lock into resonances with $\Delta\Phi = 0^\circ$ [lighter
  (green) points in Fig.~\ref{PNtidesiso}].  Once the binaries are
trapped near resonances, they evolve toward the diagonal in the
$(\theta_1,\theta_2)$ plane, as seen in the left panel of
Fig.~\ref{PNtidesiso}.  This corresponds to $\theta_{12} \to 0^\circ$
for binaries near the $\Delta\Phi = 0^\circ$ family of resonances (RMR
scenario).  As seen in the right panel of Fig.~\ref{PNtidesiso}, there
is a much broader range of final values for $\theta_{12}$ in the SMR
scenario, because these final values depend on the initial
astrophysical distribution of $\mathbf{S_0}\cdot \mathbf{\hat L}$
according to Eq.~(\ref{constrainth12}).

Fig.~\ref{PNtidespolar} shows that spin-orbit resonances can have an
even stronger effect on BH binaries when SN kicks are polar (aligned
within $\theta_b = 10^\circ$ of the stellar spin
\cite{psr-kick-Crab-Kaplan2008}).  As discussed in
Appendix~\ref{SNkicks}, exactly polar kicks tilt the orbital plane by
an angle $\Theta$ given by Eq.~(\ref{E:poltilt}), which can only
attain a maximum value $\cos^{-1} (2\beta)^{-1/2}$ (where
$\beta=M_f/M_i$ is the ratio of the total binary mass before and after
the SN) without unbinding the binary.  For $\beta \simeq 0.9$, as in
our SMR and RMR scenarios, $\Theta \lesssim 40^\circ$, and kicks are
rarely large enough even to saturate this limit.  This explains the
much narrower distribution of initial values of $\theta_i$ in the left
panel of Fig.~\ref{PNtidespolar} compared to Fig.~\ref{PNtidesiso}.
Binaries with these smaller initial misalignments are more easily
captured into resonances, as can be seen from the near total
segregation of the SMR and RMR populations in $\Delta \Phi$ by the
time the binaries reach $a_{\rm PNf} = 10M$ in the right panel of
Fig.~\ref{PNtidespolar}.

In our model, two physical mechanisms are responsible for changing BH
spin orientations: SN kicks and tidal alignment.  Both mechanisms are
critical: kicks generate misalignments between the spins and the
orbital angular momentum, but only tides can introduce the asymmetry
between these misalignments that causes one family of spin-orbit
resonances (the $\Delta\Phi = \pm 180^\circ$ family in the SMR
scenario, the $\Delta\Phi = 0^\circ$ family in the RMR scenario) to be
favored over the other. When tidal effects are removed, as shown in
Figs.~\ref{PNnotidesiso} and \ref{PNnotidespolar}, BH binaries are
formed with $\theta_1\simeq \theta_2$ on average. Being symmetric
under exchange of the two BHs, the evolution in the SMR and RMR
scenarios is almost identical.  As expected, the binaries do not lock
into resonant configurations, instead precessing freely during the
whole inspiral.  In the late stages of inspiral, the binaries tend to
pile up at $\Delta\Phi=\pm 90^\circ$, i.e. they spend more time in
configurations where the projections of the two spins on the orbital
plane are orthogonal to each other.  Unlike the spin-orbit resonances,
configurations with $\Delta\Phi=\pm 90^\circ$ are not steady-state
solutions to the spin-evolution equations in the absence of radiation
reaction \cite{2004PhRvD..70l4020S}.  The pile up at these
configurations however is an essential complement to the spin-orbit
resonances for preserving the well known result that initially
isotropic spin distributions remain isotropic (see
e.g.~\cite{2007ApJ...661L.147B}).  The physical origin of this
phenomenon merits further investigation.

\section{Comparison with population synthesis}
\label{sec:CompareToPopsyn}

We have demonstrated that viable astrophysical formation channels can
result in BH binaries that are strongly affected by spin-orbit
resonances during the late PN portion of the inspiral but before the
binary enters the GW detection band. Therefore PN resonances can
affect the observed dynamics of precessing binaries. Even more
interestingly, the distribution of the angles $\Delta \Phi$ and
$\theta_{12}$ is a diagnostic tool to constrain some of the main
physical mechanisms responsible for BH binary formation (namely the
efficiency of tides, and whether mass transfer can produce mass-ratio
reversal).

However, some caveats are in order. Even our limited exploration of
the parameter space of BH binary formation models has shown that the
influence of PN resonances depends sensitively on highly uncertain
factors, such as the magnitude and direction of SN kicks, or the mass
ratio and semimajor axis of the binary at various stages of its
evolution.  In this Section, we argue that: (i) our fiducial scenarios
are indeed representative of the predictions of more sophisticated
population-synthesis models (Section~\ref{SS:rep}); and (ii) as a
consequence, observations of spin-orbit resonances through their GW
signatures can provide valuable insight into BH binary formation
channels (Section~\ref{SS:learn}).

\subsection{Is our fiducial scenario representative?} \label{SS:rep}

In our study we chose to follow the evolution of two binary
progenitors in detail, using a specific formation channel.  The
resulting BH binaries resemble at least qualitatively the low-mass BH
binaries that can be formed through a wide range of compact-object
formation scenarios at a range of metallicities: see
e.g.~\cite{2012ApJ...759...52D}.

An important assumption made in this study is that of negligible mass
loss. Current calculations suggest that the progenitors of the most
commonly detected BH binaries will in fact have low metallicity and
strongly suppressed mass loss \cite{2012ApJ...759...52D}.
The advantage of our approach is that by neglecting mass loss and
focusing on a pair of fiducial binaries we can perform a ``controlled
experiment'' to highlight how different physical phenomena influence
the efficiency of PN resonance locking.
Variations in the range of initial binary masses, wind mass loss and
other mass transfer modes will affect the mass distribution of the
binaries and the initial distribution of the misalignment angles
$(\theta_1\,,\theta_2)$, but not our main qualitative predictions,
that should be rather robust.

This study included what we believe to be the most important physical
mechanisms that could trap binaries in resonant configurations, but it
is certainly possible that additional ingredients overlooked in our
model could complicate our simple interpretation of the results.
For example, our argument relies on a universal and deterministic
relationship between stellar masses and compact remnants.  By
contrast, some studies suggest that the relationship between the
initial and final mass may depend sensitively on interior structure
\cite{2012ApJ...757...69U}, rotation, or conceivably even
stochastically on the specific turbulent realization just prior to
explosion. As a concrete example, recent simulations of
solar-metallicity SN explosions by Ugliano et
al. \cite{2012ApJ...757...69U} (including fallback) and O'Connor and
Ott \cite{2011ApJ...730...70O} (neglecting fallback) have produced
non-monotonic relationships between the progenitor and final BH
masses.  Likewise, our argument makes the sensible assumption that BH
spins are aligned with the spin of their stellar progenitor, but
neutron star observations suggest that the protoneutron star's spin
axis may be perturbed in a SN \cite{2011ApJ...742...81F}.

Our case studies of binary evolution omit by construction many of the
complexities present in more fully developed population-synthesis
models. The inclusion of additional physics presents interesting
opportunities for a more detailed understanding of the connection
between poorly constrained assumptions in population-synthesis models
and GW observations. Some of the limitations we imposed on our model
-- and therefore, interesting opportunities for follow-up studies --
are listed below: (1) we follow the formation and evolution of only
two progenitor binaries, rather than monitoring a distribution of
masses; (2) we only consider maximally spinning BHs, while we should
consider astrophysically motivated spin magnitude distributions; (3)
we adopt very simple prescriptions for mass transfer and evolution,
which have minimal feedback onto the structure and evolution of each
star; (4) we employ an extreme ``all or nothing'' limit for tidal
interactions; (5) we assume that BHs are kicked with a specific
fraction of the overall SN kick strength; (6) we neglect stellar mass
loss, magnetic braking and other phenomena that can occur in different
formation scenarios.

In summary: while our fiducial scenario provides a representative
environment to explore the physics of PN resonances, the specific mass
distribution and the quantitative distribution of the misalignment
angles at the beginning of the PN-driven inspiral will depend on
detailed binary-evolution physics which is neglected by construction
in our toy model. It will be interesting to initialize our Monte Carlo
simulations using more comprehensive binary-evolution models that
include a distribution of progenitor masses, track tidal backreaction
on the spins and orbit, and model in more detail mass transfer and the
modifications it introduces to core and stellar evolution.

\subsection{Observational payoff} 
\label{SS:learn}

Let us provide a specific example to illustrate these uncertainties
and their potential observational payoff. Our fiducial model assumed
relatively low-mass BHs. These systems receive strong SN kicks (due to
small fallback) and are more significantly influenced by CE
contraction (because of the greater relative effect of the envelope
binding energy). By contrast, more massive BHs in the
\texttt{StarTrack} sample will accrete a significantly higher fraction
of their pre-SN mass, which drastically suppresses the typical kick
magnitude. As a result, massive BH binaries can be expected to have BH
spins more aligned with the orbital angular momentum.

This sort of qualitative difference between low- and high-mass BH
binaries presents an opportunity for GW detectors. The most easily
measurable quantity in GW observations is the ``chirp mass'' $M_{\rm
  chirp}=\eta^{3/5}M$, where $M=m_1+m_2$ is the total binary mass and
$\eta=m_1 m_2/M^2$ is the symmetric mass ratio (see
e.g. \cite{Cutler:1994ys,Poisson:1995ef}). Therefore, even though
current simulations suggest that the detected sample will be dominated
by high-mass, nearly aligned BH binaries, observations can clearly
identify the low-mass sample, which should exhibit significant initial
misalignment and more interesting precessional dynamics.
Given the significant uncertainties in population-synthesis models,
even upper limits on the spin-orbit misalignment for high-mass BH
binaries would be extremely valuable, either to corroborate the
expectation of strong alignment or to demonstrate the significance of
SN kicks for high-mass BHs.

\begin{figure}[thb]
\includegraphics[width=\columnwidth]{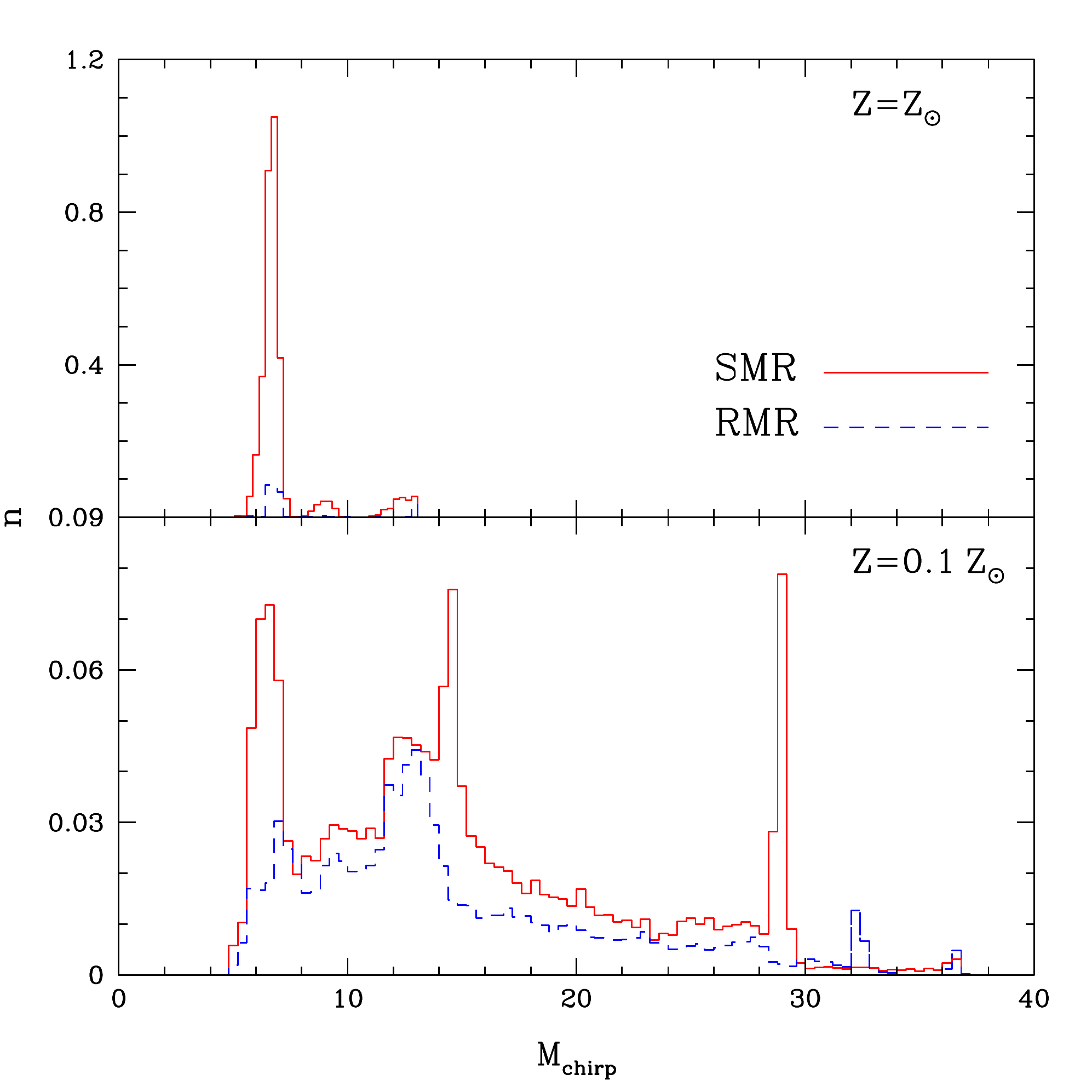}
\caption{Histograms of binaries that do (RMR) or do not (SMR) undergo
  mass-ratio reversal as a function of chirp mass according to the
  publicly available \texttt{StarTrack} data from
  \url{http://www.syntheticuniverse.org/}. For illustration, here we
  choose Subvariation A of the standard model, in the terminology of
  \cite{2012ApJ...759...52D}. A comparison of the upper and lower
  panels shows the striking differences in the chirp-mass distribution
  resulting from different choices for the metallicity $Z$.
\label{fig:RatioVersusMass}
}
\end{figure}

\begin{table*}[thb]
\begin{tabular}{|ll|rrr|rrr|rrr|rrr|}
\hline\hline
&Variation
&\multicolumn{3}{c|}{Subvariation A}
&\multicolumn{3}{c|}{Subvariation B}
&\multicolumn{3}{c|}{Subvariation A} 
&\multicolumn{3}{c|}{Subvariation B}  \\
&
&\multicolumn{3}{c|}{$Z/Z_\odot=0.1$} 
&\multicolumn{3}{c|}{$Z/Z_\odot=0.1$}
&\multicolumn{3}{c|}{$Z/Z_\odot=1$} 
&\multicolumn{3}{c|}{$Z/Z_\odot=1$}  \\
&& SMR $\:$& RMR$\:$ & \#&  SMR$\:$ & RMR$\:$ & \#&  SMR$\:$ & RMR$\:$ & \#&  SMR$\:$ & RMR$\:$ & \#\\
\hline\hline
0: 	& Standard			&63.2\%	&36.8\%	&32496	 &66.8\% &33.2\%	&17038	&91.9\%	&8.1\%	&10160	 &92.9\%	&7.1\%	&8795 \\ 
1: 	&$\lambda=0.01$ 		&67.9\%	&32.1\%	&12368	 &67.4\% &32.6\%	&11401	&93.6\%	&6.4\%	&8171	 &93.6\%	&6.4\%	&8171 \\  
2: 	&$\lambda=0.1$			&62.7\%	&37.3\%	&27698	 &65.2\% &34.8\%	&16885	&88.9\%	&11.1\%	&11977	 &92.1\%	&7.9\%	&8577 \\ 
3: 	&$\lambda=1$			&54.2\%	&45.8\%	&51806	 &65.7\% &34.3\%	&19415	&79.1\%	&20.9\%	&15820	 &91.6\%	&8.4\%	&8442 \\ 
4: 	&$\lambda=10$			&50.1\%	&49.9\%	&50884	 &62.9\% &37.1\%	&17939	&73.2\%	&26.8\%	&14425	 &91.6\%	&8.4\%	&8321 \\ 
5: 	&$M_{\rm NS}=3 M_\sun$		&62.5\%	&37.5\%	&32236	 &66.2\% &33.8\%	&16868	&91.6\%	&8.4\%	&9972	 &92.8\%	&7.2\%	&8589 \\ 
6: 	&$M_{\rm NS}=2 M_\sun$		&62.3\%	&37.7\%	&32535	 &65.9\% &34.1\%	&16804	&91.5\%	&8.5\%	&9922	 &92.5\%	&7.5\%	&8590 \\ 
7: 	&$\sigma=132.5\text{km/s}$	&58.2\%	&41.8\%	&36546	 &63.1\% &36.9\%	&18935	&88.9\%	&11.1\%	&11099	 &89.6\%	&10.4\%	&9334 \\  
8: 	&$v_{\rm BH}=v_{\rm pNS}$ (BHs)		&56.2\%	&43.8\%	&948	 &72.5\% &27.5\%	&207	&56.2\%	&43.8\%	&16	 &0\%	        &100\%	&2    \\ 
9: 	&$v_{\rm BH}=0$ (BHs)			&56.3\%	&43.7\%	&52832	 &58.8\% &41.2\%	&34569	&66.3\%	&33.7\%	&35267	 &65.2\%	&34.8\%	&32547\\ 
10: 	& Delayed SN			&61.4\%	&38.6\%	&27310	 &66.3\% &33.7\%	&13841	&81.5\%	&18.5\%	&1032	 &81.2\%	&18.8\%	&881  \\ 
11: 	&Weak winds			&58.4\%	&41.6\%	&33872	 &63.6\% &36.4\%	&17765	&70.5\%	&29.5\%	&21786	 &64.2\%	&35.8\%	&16182\\  
\hline\hline
\end{tabular}
\caption{BH binary rates predicted by \texttt{StarTrack}. RMR (SMR) is
  the percentage of binaries that do (not) experience mass-ratio
  reversal due to mass transfer; $\#$ indicates the total number of BH
  binaries in the sample. Each row refers to a different variation
  over the ``standard model''. The variations illustrate the effect of
  changing one parameter (CE binding energy $\lambda$, kick magnitude,
  etc.) with respect to the ``best guesses'' of the standard
  model. Each row also shows the effect of changing the metallicity
  $Z$ and the Hertzsprung-gap donor prescription. In Subvariation A
  (B), binaries can (can not) survive a common-envelope event during
  the Hertzprung-gap phase; see \cite{2012ApJ...759...52D} for
  details).}
  
\label{dominiktable}
\end{table*}

Based on our prototype study, let us assume that each PN resonance is
an unambiguous indicator of a specific fomation scenario: hypothetical
GW measurements of angles $\Delta \Phi\sim \pm 180^\circ$ mean
efficient tides in the ``standard mass ratio'' (SMR) scenario;
measurements of $\Delta \Phi\sim 0^\circ$ mean that mass reversal also
occurred (RMR); finally, $\Delta \Phi\sim \pm 90^\circ$ is an
indication that tidal effects were inefficient
(cf. Fig.~\ref{scheme}). Under these assumptions, statistically
significant measurements of $\Delta \Phi$ could directly identify how
often each of the three formation channels (efficient tides, SMR;
efficient tides, RMR; inefficient tides) occurs, for each binary mass.

To illustrate how informative these measurements might be,
Fig.~\ref{fig:RatioVersusMass} shows the relative number of merging
binaries that undergo mass-ratio reversal as a function of chirp mass,
as derived from the most recent \texttt{StarTrack} binary-evolution
models \cite{2012ApJ...759...52D}.
The figure (which is meant to be purely illustrative) refers to
Subvariation A of the ``standard model'' of Dominik et al.
\cite{2012ApJ...759...52D}. Each panel shows the chirp-mass
distribution of binaries that either do (RMR, dashed blue histograms)
or do not (SMR, red solid histograms) undergo mass-ratio
reversal. This distribution has characteristic ``peaks'' at specific
values of the chirp mass at any given $Z$ and it depends very strongly
on composition, as we can see by comparing the two panels (which refer
to $Z/Z_\odot=1$ and $Z/Z_\odot=0.1$, respectively). According to our
model, measurements of $\Delta \Phi$ for a large enough sample of
binaries would allow us to reconstruct the shape of these histograms
as a function of chirp mass, potentially enabling new high-precision
tests of binary evolution, above and beyond the information provided
by the mass distribution alone.

A preliminary assessment of the main features of population-synthesis
models that could be probed by these measurements can be inferred from
Table \ref{dominiktable}. There we list the overall fraction of BH
binary systems that undergo mass-ratio reversal for several different
binary-evolution scenarios explored in \cite{2012ApJ...759...52D}.
The most dramatic difference is due to composition: with few
exceptions, models with solar composition ($Z/Z_\odot=1$) almost
exclusively produce SMR binaries, while models with subsolar
composition ($Z/Z_\odot=0.1$) produce comparable proportions of SMR
and RMR binaries.
Furthermore there are clear trends in the ratio RMR/SMR as a function
of the envelope-binding-energy parameter $\lambda$ discussed in
Appendix~\ref{A:CE} (compare variations 1 to 4); the strength of SN
kicks (variations 8 and 9); and the amount of mass loss through winds
(variation 11).  These parameters are also well known to significantly
influence the overall number and mass distribution of merging
binaries.

In conclusion, while our model needs further testing and scrutiny
against more complete population-synthesis calculations, it strongly
indicates that GW measurements of $\Delta \Phi$ and $\theta_{12}$ will
provide a useful diagnostic of compact binary formation, complementary
to the more familiar mass and spin measurements. In the next Section
we conclude the paper with an overview of the challenges and rewards
associated with these measurements.

\section{Discussion} \label{results}

Previous Monte Carlo studies of the spin-orbit resonances discovered
by Schnittman \cite{2004PhRvD..70l4020S} showed that spins tend to
lock in a resonant plane if the binary has mass ratio $q \gtrsim 0.4$
and the dimensionless spin magnitudes $\chi_i \gtrsim 0.5$ as long as
there is an initial asymmetry in the relative orientation of the spins
with respect to the orbital angular momentum, i.e. $\theta_1\neq
\theta_2$
\cite{2010PhRvD..81h4054K,2010ApJ...715.1006K,2012PhRvD..85l4049B}.

In this work we built a toy model for BH binary formation focusing on
the main physical ingredients that can produce such an asymmetry: SN
kicks (that tilt the orbital plane every time a BH is formed), tidal
interactions (that tend to realign the spin of the star that collapses
later with the orbital angular momentum) and mass transfer (that can
produce mass-ratio reversal, so that the heaviest BH corresponds to
the lighter stellar progenitor).  We showed that for stellar-mass
compact objects formed at the endpoint of isolated binary evolution
the required conditions should ubiquitously occur.

Perhaps more interestingly, we demonstrated that the angle
$\Delta\Phi$ between the components of the BH spins in the plane
orthogonal to the orbital angular momentum is in one-to-one
correspondence with the BH formation channel that gave birth to the BH
binary: if tides are efficient the PN evolution attracts the spins to
the resonant plane with $\Delta\Phi \simeq 0^\circ$ ($\Delta\Phi
\simeq \pm 180^\circ$) if mass reversal does (does not) occur.  When
tidal effects are inefficient the spins precess freely, and they pile
up at $\Delta\Phi=\pm 90^\circ$ by the time the binary enters the band
of advanced GW detectors.
A preliminary comparison with detailed population-synthesis
calculations suggests that the fraction of binaries in each family of
resonant configurations, both overall and as a function of (chirp)
mass, should provide a highly informative diagnostic on some of the
main uncertainties involved in binary-evolution physics (metallicity,
binding energy of the CE, magnitude of BH kicks).  Measuring this
fraction will require a large sample of BH mergers with sufficient
signal-to-noise ratio, but hopefully such a sample will be obtainable
by Advanced LIGO/Virgo after some years of operation at design
sensitivity.

Our initial study merits detailed follow-ups to assess (i) the
potential accuracy of GW measurements of the precessional parameters,
and (ii) the information that can be extracted by comparison with
population-synthesis models.

Detailed studies are required from the point of view of GW data
analysis. We have assumed for simplicity that each PN resonance can be
easily and unambigously distinguished. In practice, accurate
matched-filtering measurements of the angles $\Delta \Phi$ and
$\theta_{12}$ will need more work on the GW source-modeling
front. Relevant issues here include the construction of
gravitational-waveform templates adapted to resonant configurations,
the development of specialized parameter-estimation strategies and the
understanding of systematic (as opposed to statistical) errors for
second- and third-generation detectors. Spin modulations are known to
influence both the amplitude and phase of the emitted radiation, and
while there are several preliminary investigations of parameter
estimation from spinning, precessing binaries, the direct measurement
of parameters characterizing the spin-orbit resonances may require the
inclusion of higher-order spin terms and/or higher harmonics in the
waveform models.

From an astrophysical standpoint, the observable distribution of
binary systems as they enter the detector band should be calculated
(more realistically) by applying our PN evolution to initial data
derived from state-of-the-art binary population-synthesis models.  In
addition to corroborating our results, such a study will establish a
comprehensive library of reference models that can be compared to
observational data using Bayesian or other model-selection strategies:
see e.g.
\cite{2010CQGra..27k4007M,O'Shaughnessy:2006wh,O'Shaughnessy:2009ft,Mandel:2009pc,Sesana:2010wy,Gair:2010bx,O'Shaughnessy:2012zc,Sesana:2010wy,Gair:2010bx}
for previous efforts in this direction. Such a study is necessary also
to make contact with other observables, such as as the rate and mass
distribution of compact binaries. Only with a comprehensive and
self-consistent set of predictions can we quantify how much the
information provided by PN resonances complements information
available through other observable quantities.

In conclusion, the direct observation of resonant locking will be
challenging from a GW data-analysis standpoint. However the
relatively transparent astrophysical interpretation of PN resonances
makes such an investigation worthwhile. Even if only observationally
accessible for the loudest signals, these resonances will enable
unique insights into the evolutionary channels that produce merging
compact binaries. In our opinion, more detailed studies of resonant
locking in connection with population-synthesis models will offer a
great observational opportunity for GW astronomy.

\section*{Acknowledgments}

We are grateful to Parameswaran Ajith, Chris Belczynski, Tomasz Bulik,
Marco Cavagli\`a, Marc Favata, Or Graur, Michael Horbatsch, Giuseppe
Lodato and Sterl Phinney for useful discussions and suggestions on
various aspects of this work. This research was supported in part by
NSF Grant No.~PHY11-25915 and by the LIGO REU program at the
California Institute of Technology. ROS was supported by NSF Grant
No.~PHY-0970074. EB and DG were supported by NSF CAREER Grant
No.~PHY-1055103.  EB and US acknowledge support from
FP7-PEOPLE-2011-IRSES Grant No.~NRHEP-295189 and NSF-XSEDE Grant
No.~PHY-090003.
US also acknowledges support from
FP7-PEOPLE-2011-CIG Grant No.~CBHEO-293412,
CESGA Grant No.~ICTS-234,
BSC, RES Grant No.~AECT-2012-3-0011,
ERC Starting Grant No.~DyBHo 256667
and the Cosmos system, part of DiRAC, funded by STFC and BIS.

\appendix
\section{Binary-evolution phenomenology}
\label{ap:BinEv}

Binary population synthesis relies on copious guidance from both
observations and theory \cite{2006epbm.book.....E}.  Simulations of
binary evolution that self-consistently account for stellar structure
and mass transfer are computationally expensive and depend on a wide
variety of parameters
\cite{2006epbm.book.....E,starev-MESA-Paxton2010}.  Models that hope
to generate astrophysically realistic binary populations must tabulate
the results of these simulations and calibrate them against
observations
\cite{2006epbm.book.....E,2002MNRAS.329..897H,2008ApJS..174..223B}.
Well developed algorithms exist to quickly generate large synthetic
compact-binary populations similar to those produced in more expensive
direct simulations \cite{2002MNRAS.329..897H,2008ApJS..174..223B}.  In
this Appendix, we use such population-synthesis models to justify and
put into context the simple procedure adopted in this paper. To
further validate our model, we have also performed a handful of
detailed binary-evolution calculations with the binary-stellar
evolution \texttt{BSE} code by Hurley et al.
\cite{2002MNRAS.329..897H}.  When adopting similar assumptions (i.e.,
low stellar mass-loss rates and large envelope binding energies), the
\texttt{BSE} code produces qualitatively similar evolutionary
scenarios to the procedure outlined in the text.  The simple model and
fiducial scenarios considered in this paper do not account for a
thorough exploration of the parameter space, but they illustrate the
essential physics and demonstrate that PN resonance locking can be the
preferred outcome of astrophysically motivated BH binary formation
channels.

\subsection{Single stellar evolution} 
\label{A:InitMass}

In this Section, we provide relevant information about the evolution
of isolated stars.  Main-sequence stars born with a mass $M_S$ have a
radius \cite{1991Ap&amp;SS.181..313D}
\begin{equation} \label{E:RS}
\frac{R_S}{R_\odot}\simeq 1.33 \, \left(\frac{M_S}{M_\odot}\right)^{0.555}~.
\end{equation}
Massive, metal-rich main-sequence stars lose a substantial amount of
mass via winds prior to going SN, but we neglect this mass loss for
simplicity.  The inclusion of wind mass loss in our model would reduce
the mass of the hydrogen envelope available to be transferred to the
secondary during the first mass-transfer event.  While neglecting this
mass loss quantitatively changes the binary evolution, we believe that
it does not qualitatively alter our conclusions.  Larger (and
appropriately chosen) initial stellar masses would lead to final BH
binaries with masses comparable to those considered in our model even
in the presence of winds.

Stars with main-sequence masses in the range $25~M_\odot \leq M_S \leq
40~M_\odot$ evolve into supergiants with helium-core masses
well approximated by
\begin{equation} \label{E:MC}
M_C \simeq 0.1 M_S + 5 M_\sun
\end{equation}
(cf.~top panel of Fig.~14 of \cite{2008ApJS..174..223B}) and radii
\cite{1995MNRAS.273..731R}
\begin{equation} \label{E:RG}
\frac{R_G}{R_\odot} \simeq 4950 \frac{(M_C/M_\odot)^{4.5}}{1+4 (M_C/M_\odot)^{4}}+0.5~.
\end{equation}
Once the hydrogen envelopes have been lost, the naked helium cores
have radii \cite{1998ApJ...502L...9F}
\begin{equation} \label{E:RC}
\log\frac{R_C}{R_\odot} \simeq -0.699 + 0.0557 \left( \log\frac{M_C}{M_\odot} - 0.172\right)^{-2.5}~.
\end{equation}
We neglect further evolution of the naked helium star before SN.  For
the large masses typical of BH progenitors, the naked helium cores
have radiative envelopes and do not expand substantially during
subsequent shell burning
\cite{2003ApJ...592..475I,2008ApJS..174..223B}.  After going SN, a
main-sequence star leaves behind a BH of mass (bottom panel of Fig.~14
of \cite{2008ApJS..174..223B})
\begin{equation} \label{E:MBH}
M_{\rm BH} \simeq 0.3 M_S - 3 M_\sun~.
\end{equation}

\subsection{Initial semimajor axis} 
\label{A:InitSep}

The initial binary separation $a_0$ is drawn from a uniform
logarithmic distribution in the range [$a_{\rm min}$, $a_{\rm max}$]
\cite{2008ApJS..174..223B,1924PTarO..25f...1O,
  1983ARA&A..21..343A,2007IAUS..240..417P}.  The upper limit $a_{\rm
  max}$ is chosen to ensure that the primary fills its Roche lobe
during its supergiant phase, while the lower limit $a_{\rm min}$ is
chosen so that the secondary does {\it not} fill its Roche lobe after
receiving mass from the primary.  The Roche-lobe radius $R_L$ of a
star of mass $m_\alpha$ in an orbit of semimajor axis $a$ about a
companion of mass $m_\beta$ is
\cite{1983ApJ...268..368E,2002MNRAS.329..897H}
\begin{equation} \label{E:RLdef}
R_L(a, m_\alpha, m_\beta) \simeq \frac{0.49Q^{2/3}}{0.6Q^{2/3} + \ln (1 + Q^{1/3})}a~,
\end{equation}
where $Q \equiv m_\alpha/m_\beta$, so the above limits are determined
by the constraints
\begin{align}
R_L(a_{\rm max}, M'_{Si}, M''_{Si}) = R'_G,
\\
R_L(a_{\rm min}, M''_{Sf}, M'_C) = R''_{Sf}.
\end{align}
These limits are somewhat arbitrary, but different choices would not
affect our main results. In fact, binaries that do not go through mass
transfer ($a > a_{\rm max}$) are so widely separated that they are
easily unbound by the first SN, while binaries where mass is
transferred back to the primary prior to this SN ($a < a_{\rm min}$)
will merge in the CE phase.  These limits will therefore only affect
the failure fractions presented in Table~\ref{rates}, not the spin
alignments of merging BH binaries.

\subsection{Stable mass transfer} 
\label{A:MT}

When a star fills its Roche lobe, gas will either be stably
transferred to its companion or form a CE about both members of the
binary.  Stable mass transfer is discussed in this Section of the
Appendix, while CE evolution is discussed in Section~\ref{A:CE}.  In
general, the stability of mass transfer depends on the donor star, the
accreting star, and the mass ejected to infinity; as a first
approximation, stability criteria are usually implemented by simple
thresholds on the binary mass ratio, as summarized in
\cite{2012ApJ...746..186C} and references therein.  For our mass
ratios, mass transfer from the primary to the secondary prior to the
first SN will be stable, while mass transfer from the secondary to the
primary between the two SN events will lead to the formation of a CE.
A fraction $f_a$ of the mass lost by the primary in the first
mass-transfer event will be accreted by the secondary, increasing its
mass to
\begin{equation}
\label{fa_def}
M''_{Sf} = M''_{Si} + f_a(M'_{Si}-M'_C)~.
\end{equation}
Fully conservative mass transfer ($f_a = 1$) preserves the total mass
of the system, while all of the mass lost by the donor is ejected from
the system in fully non-conservative mass transfer ($f_a = 0$).  We
assume that stable mass transfer is semiconservative ($f_a = 1/2$), in
agreement with the standard model of Dominik et
al. \cite{2012ApJ...759...52D}. Larger values of $f_a$ during this
first mass-transfer event will tend to favor the RMR scenario over the
SMR scenario.  Since $f_a$ is directly tied to the fraction of
binaries that undergo mass-ratio reversal in a given mass and
mass-ratio range, our model suggests that it is potentially measurable
via GW observations. For simplicity, we assume that tides and the mass
transfer itself efficiently circularize the orbit (but see
\cite{2009ApJ...702.1387S,2010ApJ...724..546S} for recent
investigations of mass transfer and circularization in eccentric
binaries).

\subsection{Supernova kicks: magnitude and direction} 
\label{A:kick}

Following \cite{2008ApJ...682..474B}, we assume that asymmetric SN
events impart hydrodynamical recoils to the newly formed protoneutron
stars.  We calibrate the magnitude of this primordial kick using
observed proper motions of young pulsars: each protoneutron star is
kicked with a velocity $v_{\rm pNS}$ drawn from a single Maxwellian
with parameter $\sigma=265$ km/s \cite{2005MNRAS.360..974H}.  A
fraction $f_{\rm fb}$ of this asymmetrically ejected material falls back
onto the protoneutron star and is accreted as it collapses into a BH.
This fallback suppresses the magnitude of the final kick imparted to
the BH to $v_{\rm BH} \simeq (1 - f_{\rm fb}) v_{\rm pNS}$; for BHs with
masses $M_{\rm BH} = (6 M_\odot, 7.5 M_\odot)$, as in our fiducial
scenarios, simulations suggest $f_{\rm fb} \simeq 0.8$
\cite{1999ApJ...522..413F,2001ApJ...554..548F}.  This BH kick
distribution is consistent with the observed proper motions of
galactic X-ray binaries hosting BHs
\cite{2009ApJ...697.1057F,2012ApJ...747..111W}.  Although our results
are not extremely sensitive to the precise magnitude of the BH kicks,
the existence of such kicks is crucial to our model, as they are the
only observationally well motivated mechanism to introduce
misalignment between the compact binary spins and the orbital plane.

We assume that the BH kicks are distributed in a double cone of
opening angle $\theta_b$ about the BH spin and consider two extreme
scenarios: isotropic ($\theta_b=90^\circ$) or polar
($\theta_b=10^\circ$) kicks. There is some observational
\cite{2006ApJ...639.1007W,arXiv:1301.1265} and theoretical
\cite{1998Natur.393..139S,2001ApJ...549.1111L} support for the polar
model. However we examine both possibilities because this choice has a
significant effect on the resulting binary orbits, as discussed in
Appendix~\ref{SNkicks} below.  Our choice of $\theta_b=10^\circ$ in the
polar model was partly motivated by a comparable observed misalignment
between the spin and proper motion of the Crab pulsar
\cite{psr-kick-Crab-Kaplan2008}.
 
\subsection{Supernova kicks: influence on the orbit}
\label{SNkicks} 

\begin{figure}\centering
\includegraphics[width=0.48\textwidth]{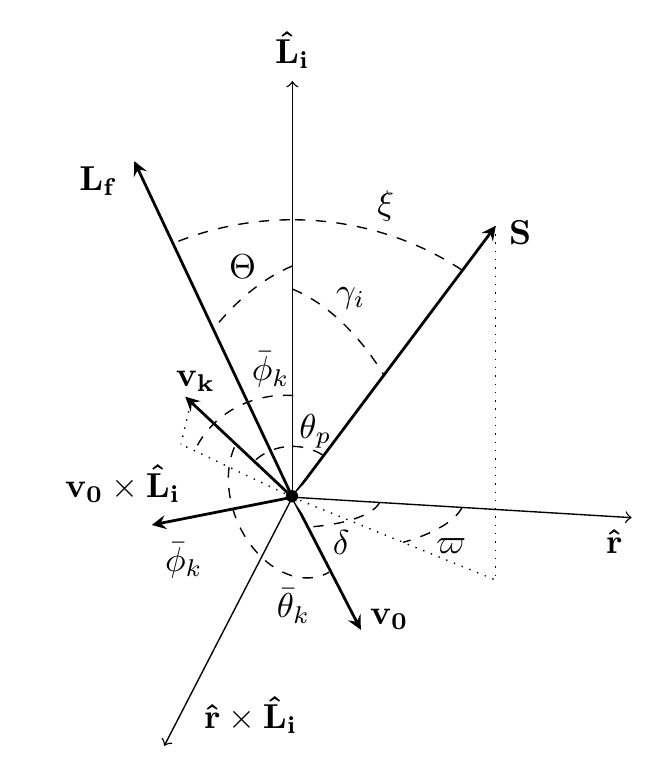}
\caption{Definitions of the angles used in Appendix \ref{SNkicks} to
  study SN kicks.  Before the SN, the members of the binary have a
  separation $\mathbf{r}$ and relative velocity $\mathbf{v_0}$.  Mass
  loss and the SN kick $\mathbf{v_k}$ tilt the orbital angular
  momentum from $\mathbf{L_i}$ to $\mathbf{L_f}$ while leaving the
  spin $\mathbf{S}$ unchanged.}
\label{refsys2}
\end{figure}

In this Section, we describe how SN kicks are implemented in our Monte
Carlo calculations. The expressions provided below have been published
previously either under more restrictive assumptions
\cite{2000ApJ...541..319K} or using different notation
\cite{2002MNRAS.329..897H}, but we rederive them here for clarity and
completeness.  Each SN reduces the mass of the binary and imparts a
kick to the newly produced compact remnant.  We calculate how these
effects change the Keplerian orbital elements by applying energy and
angular-momentum conservation to the binary before and after the SN.
As the duration of the SN explosion is short compared to the other
stages of binary evolution, we assume that this orbital modification
occurs instantaneously. The definitions of the angles used in this
Appendix are illustrated in Fig.~\ref{refsys2}.

In the simulations reported in this paper we assume that the binary is
on a circular orbit ($e_i=0$) and that the stellar spins are aligned
with the orbital angular momentum ($\gamma_i=0$) when the first SN
occurs (we have actually relaxed the circularity assumption in
additional simulations not presented here, and we verified that this
has a negligible impact on our conclusions).  If tides are
inefficient, both of these simplifying assumptions will not hold, in
general, for the second SN. Therefore here we present general
expressions for the post-SN orbital elements.  These expression were
first derived (to our knowledge) in \cite{2002MNRAS.329..897H}, but
here we use notation similar to that of Kalogera
\cite{2000ApJ...541..319K}.

The binary separation $r$ for a Keplerian orbit with initial
semimajor axis $a_i$ and eccentricity $e_i$ can be expressed as
\begin{align}
r=\frac{a_i (1-e_i^2)}{1+e_i\cos\psi_i},
\label{ellipse}
\end{align}
where $\psi_i$ is the true anomaly.  Values for the true anomaly at
the time of the SN are chosen by assuming that the explosion is
equally likely to occur at any given time.  The time $t$ after the
binary reaches pericenter is given by
\begin{align}
\frac{2 \pi}{P} t = E - e_i \sin E,
\label{kepeq}
\end{align}
where
\begin{align}
P = 2\pi \left( \frac{a_i^3}{GM_i} \right)^{1/2}
\end{align}
is the period of a binary of total mass $M_i$.  The eccentric anomaly
$E$ is related to the true anomaly $\psi_i$ by
\begin{align}
\cos\psi_i=\frac{\cos E - e_i}{ 1- e_i \cos E}~.
\end{align}
We assume that $t$ is uniformly distributed in the range $[0,P]$ and
derive the corresponding values of $\psi_i$ from these relations.

The direction of the kick velocity $\mathbf{v_k}$ is defined by a
polar angle $\bar{\theta}_k$ and an azimuthal angle $\bar{\phi}_k$.
Here $\bar{\theta}_k$ is the angle between $\mathbf{v_k}$ and the
pre-SN orbital velocity $\mathbf{v_0}$, and the axis defined by
$\bar{\phi}_k = 0$ is chosen to be parallel to the orbital angular
momentum $\mathbf{L}$ (see Fig.~\ref{refsys2}).  The direction of the
spin $\mathbf{S}$ of the collapsing star is specified by the angle
$\gamma_i$ between $\mathbf{S}$ and $\mathbf{L}$ and the angle
$\varpi$ between the projection of $\mathbf{S}$ in the orbital plane
and the separation $\hat{\mathbf{r}}$ between the members of the
binary.  In terms of these angles, the angle $\theta_p$ between
$\mathbf{S}$ and $\mathbf{v_k}$ is given by
\begin{align}
\cos\theta_p&= -\left( \sin\bar{\theta}_k \sin\bar{\phi}_k \sin \delta  + \cos\bar{\theta}_k \cos \delta \right) \cos \varpi \sin \gamma_i 
\notag \\
&+\left( \cos\bar{\theta}_k \sin \delta  - \sin\bar{\theta}_k \sin\bar{\phi}_k \cos \delta \right) \sin \varpi  \sin \gamma_i  
\notag \\
&+ \sin \bar{\theta}_k \cos \bar{\phi}_k \cos \gamma_i~,
\end{align}
where the angle $\delta$ between the orbital velocity and line of
separation is given in terms of the true anomaly by
\begin{align}
\cos\delta=\frac{e_i\sin \psi_i}{\left(1+2e_i\cos \psi_i+ e^2_i\right)^{1/2}}~.
\end{align}
In our Monte Carlo simulations, kick directions are drawn from uniform
distributions in $\bar{\phi}_k$, $\cos \bar{\theta}_k$, and $\varpi$.  Kicks
confined to within an angle $\theta_b$ of the stellar spin
$\mathbf{S}$ are therefore implemented by repeated draws from this
distribution such that
\begin{align}
\theta_p\leq\theta_b \qquad \text{or} \qquad \theta_p\geq\pi-\theta_b.
\label{polarconstraint}
\end{align}

The SN reduces the total mass of the binary from $M_i$ to $M_f$ and
changes the velocity of the exploding star from $\mathbf{v_0}$ to
$\mathbf{v_0} + \mathbf{v_k}$.  Applying energy and angular-momentum
conservation to the binary before and after the SN, we find that the
final semimajor axis $a_f$ and eccentricity $e_f$ are given by
\cite{2002MNRAS.329..897H}
\begin{align}
a_f&= a_i \,\beta \bigg[ 2\left( \beta-1\right) \frac{1+e_i \cos\psi_i}{1-e_i^2} +1-u_k^2 
\notag\\
 &- 2 u_k  \left( \frac{1+ 2e_i \cos\psi_i + e_i^2}{1-e_i^2} \right)^{1/2}\cos\bar{\theta}_k \bigg]^{-1},
\label{Ksemimajoraxis}
\end{align}

\begin{widetext}
\begin{align}
1 - e_f^2 = \frac{1- e_i^2}{\beta^2} \Bigg\{ &\Bigg[ 1 + u_k  \left( \frac{1-e_i^2}{ 1 + 2e_i \cos \psi_i +e_i^2} \right)^{1/2}
\left( \cos \bar{\theta}_k -  \frac{e_i \sin \psi_i \sin \bar{\theta}_k \sin \bar{\phi}_k}{1 + e_i \cos \psi_0} \right) \Bigg]^2 
+ (1-e_i^2) \left( \frac{u_k\sin\bar{\theta}_k \cos\bar{\phi}_k}{1+e_i \cos\psi_i}  \right)^2 \Bigg\}
\notag \\
&\times \bigg[ 2\left( \beta-1\right) \frac{1+e_i \cos\psi_i}{1-e_i^2} +1-u_k^2 
- 2 u_k  \left( \frac{1+ 2e_i \cos\psi_i + e_i^2}{1-e_i^2} \right)^{1/2}\cos\bar{\theta}_k \bigg]
,
\label{Keccentricity}
\end{align}
\end{widetext}
where $\beta=M_f/M_i$ and $u_k$ is the magnitude of the kick velocity
normalized to the circular orbital velocity before the explosion, i.e.
\begin{align}
u_k=v_k\sqrt\frac{a_i}{G M_i}~.
\end{align}
If the right-hand side of Eq.~(\ref{Keccentricity}) is negative, $e_f
> 1$ and the SN has unbound the binary.  For binaries that remain
bound, the orbital plane is tilted by an angle $\Theta$ such that

\begin{widetext}
\begin{align}
\cos \Theta &= \Bigg[ 1 + u_k  \left( \frac{1-e_i^2}{ 1 + 2e_i \cos \psi_i +e_i^2} \right)^{1/2}
\left( \cos \bar{\theta}_k -  \frac{e_i \sin \psi_i \sin \bar{\theta}_k \sin \bar{\phi}_k}{1 + e_i \cos \psi_i} \right) \Bigg] \notag \\
& \times  \Bigg\{ \Bigg[ 1 + u_k  \left( \frac{1-e_i^2}{ 1 + 2e_i \cos \psi_i +e_i^2} \right)^{1/2}
\left( \cos \bar{\theta}_k -  \frac{e_i \sin \psi_i \sin \bar{\theta}_k \sin \bar{\phi}_k}{1 + e_i \cos \psi_i} \right) \Bigg]^2 
+ (1-e_i^2) \left( \frac{u_k\sin\bar{\theta}_k \cos\bar{\phi}_k}{1+e_i \cos\psi_i}  \right)^2 \Bigg\}^{-1/2}\,,
\label{tiltangles}
\end{align}
\end{widetext}
and the angle between $\mathbf{S}$ and $\mathbf{L}$ is changed from
$\gamma_i$ to $\xi$, where
\begin{widetext}
\begin{align}
\cos \xi = &\Bigg\{\Bigg[ 1 + u_k  \left( \frac{1-e_i^2}{ 1 + 2e_i \cos \psi_i +e_i^2} \right)^{1/2}
  \left( \cos \bar{\theta}_k -  \frac{e_i \sin \psi_i \sin \bar{\theta}_k \sin \bar{\phi}_k}{1 + e_i \cos \psi_i} \right) \Bigg]
\cos\gamma_i
\notag \\
&- u_k\frac{\sqrt{1-e_i^2}}{1+e_i \cos\psi_i} \sin\bar{\theta}_k \cos\bar{\phi}_k \sin\gamma_i\sin\varpi
\Bigg\}
   \notag \\
& \times  \Bigg\{ \Bigg[ 1 + u_k  \left( \frac{1-e_i^2}{ 1 + 2e_i \cos \psi_i +e_i^2} \right)^{1/2}
 \left( \cos \bar{\theta}_k -  \frac{e_i \sin \psi_i \sin \bar{\theta}_k \sin \bar{\phi}_k}{1 + e_i \cos \psi_i} \right) \Bigg]^2 \notag \\
&+ (1-e_i^2) \left( \frac{u_k\sin\bar{\theta}_k \cos\bar{\phi}_k}{1+e_i \cos\psi_i}  \right)^2 \Bigg\}^{-1/2}.
\label{spintiltangle}
\end{align}
\end{widetext}
When $\mathbf{S}$ is aligned with $\mathbf{L}$ before the SN ($\gamma_i = 0$), the tilt of the orbital plane equals
the misalignment of the exploding star's spin ($\xi = \Theta$).

The above expressions greatly simplify for initially circular
binaries.  For example, the SN will disrupt the binary if
\begin{equation}
u_k^2+2u_k \cos \bar{\theta}_k + 1-2\beta > 0 \quad (e_i = 0)\,.
\end{equation}
The equations simplify even further if $\mathbf{S}$ and $\mathbf{L}$
are initially aligned ($\gamma_i = 0$), in which case exactly polar
kicks are given by $\bar{\theta}_k=\pi/2, \bar{\phi}_k=0$.  Exactly
polar kicks larger than $u_k > \sqrt{2\beta-1}$ always unbind the
binary, while for isotropic kicks a bound tail of the distribution
remains provided $u_k < 1 + \sqrt{2\beta}$.  If kicks are confined to
cones within an angle $\theta_b$ of $\mathbf{L}$, the minimum final
semimajor axis is
\begin{equation} \label{E:afmin}
a_{f, {\rm Min}} = \frac{a_i \beta}{2\beta - \cos^2 \theta_b} 
\quad (e_i = 0, \gamma_i = 0)~;
\end{equation}
exactly polar kicks ($\theta_b = 0$) can only increase the semimajor
axis ($a_{f, {\rm Min}} > a_1$), while isotropic kicks ($\theta_b =
90^\circ$) can reduce the semimajor axis by at most a factor of 2
($a_{f, {\rm Min}} = a_1/2$).

Exactly polar kicks also add a significant component of angular
momentum perpendicular to the initial orbital plane, leading to a
strong spin tilt:
\begin{equation} \label{E:poltilt}
\cos \Theta  = \frac{1}{\sqrt{1+u_k^2}} \quad (e_i =0, \gamma_i = 0)\,.
\end{equation}
However, the maximum tilt that polar kicks can produce while the
binary remains bound is
\begin{equation} \label{eq:TiltMax}
\Theta = \cos^{-1} (2\beta)^{-1/2}\,.
\end{equation}
By contrast, isotropic kicks can make the binary more tightly bound,
allowing greater latitude for kicks to produce bound systems with
large spin misalignments.

In the limit that the kick velocity is small compared to the orbital
velocity ($u_k \ll 1$), as should be the case for the second SN after
CE evolution has reduced the binary separation, the tilt of the
orbital plane is given by
\begin{equation} \label{E:smalltilt}
\Theta =  u_k \sin\bar{\theta}_k |\cos\bar{\phi}_k| + \mathcal{O}(u_k^{3/2}) \quad (e_i =0, \gamma_i = 0) \, .
\end{equation}

\subsection{Tidal alignment}
\label{tides}

As discussed in Sec.~\ref{SS:example}, tidal dissipation can
circularize the orbit of the binary and align the spin of the
secondary with the orbital angular momentum between the two SN
explosions
\cite{1975AA....41..329Z,1981A&amp;A....99..126H,2002MNRAS.329..897H}.
A detailed treatment of the theory of tidal damping in massive stars
is far beyond the scope of this paper, and relatively little data
exists to calibrate these theoretical models if we wished to do so.
We therefore only consider the two extreme possibilities: tides can
either fully circularize the binary and align the spin of the
secondary, or they are completely inefficient.  We provide
order-of-magnitude estimates for tidal processes below; those
interested in more details should consult one of the several excellent
published reviews of tidal processes
\cite{1981A&amp;A....99..126H,2001ApJ...562.1012E,2006epbm.book.....E}.

Tides should generally act on both members of the binary. However
tidal effects on the BH can safely be ignored, given its small size.
We therefore focus on tidal effects on the secondary between the two
SN (phase \textbf{d} of the evolutionary scenario presented in
Fig.~\ref{model}).  If the secondary is fully convective, as expected
for the core of a BH progenitor, convection causes internal damping on
the viscous timescale $t_V \simeq \gamma^{-1} (3M_S
R_{S}^2/L_S)^{1/3}$, where $M_S$, $R_S$ and $L_S$ are the mass, radius
and luminosity of the secondary, and $\gamma$ is a prefactor that
depends on details of the stellar structure
\cite{2001ApJ...562.1012E}.  The orbit evolves on the tidal-friction
timescale
\begin{eqnarray} \label{E:ttf}
t_{\rm tid}&\simeq & \tilde k \, \frac{t_V}{9}  \frac{M_{\rm S}^2}{(M_{\rm BH}+M_S)M_{\rm BH}} \left( \frac{a}{R_S} \right)^8  \notag\\
&\simeq&
 4 \times 10^{-3}\, \tilde  k \gamma\, \frac{1}{ Q (1+  Q)}\left(\frac{M_{\rm S}}{10 M_\odot}\right)^{1/3}  \left(\frac{R_{\rm S}}{10 R_\odot}\right)^{2/3}  
\notag \\
& \times &\left(\frac{L_{\rm S}}{10^4 L_\odot}\right)^{-1/3} \left(\frac{a}{R_{\rm S}}\right)^8 \text{yrs},
\label{ttidal}
\end{eqnarray}
where $M_{\rm BH}$ is the mass of the primary, $Q=M_{\rm BH}/M_S$ is
the mass ratio at this stage of the evolution, and $\tilde{k}\gamma$
is a constant of order unity depending on the internal structure of
the star \cite{1981A&amp;A....99..126H}. Though the details depend on
the initial stellar spin, tidal friction should synchronize and align
the spin of the secondary with the now circular orbit on this same
short timescale \cite{2001ApJ...562.1012E}.

The most notable feature of the tidal-friction timescale $t_{\rm tid}$
given by Eq.~(\ref{E:ttf}) is its extremely steep dependence on the
ratio $a/R_S$.  While the secondary remains on the main sequence with
a radius given by Eq.~(\ref{E:RS}), this ratio is typically 100 or
greater for binaries that avoid merging during CE evolution.  This
implies that tidal alignment occurs on timescales much longer than the
Hubble time $t_H \simeq 10^{10}$~yrs.  However, once the secondary
evolves to fill its Roche lobe, its radius is given by
Eq.~(\ref{E:RLdef}) and the ratio $a/R_S$ becomes of order unity.
This reduces the tidal-friction timescale well below typical
stellar-evolution timescales of a few million years (hydrogen-core
burning) or even the briefer time
\begin{align}
t_{\rm HG} &\simeq 2.7 \times 10^{4} \left(\frac{M_{\rm C}}{10 M_\sun}\right)^2 \left(\frac{R_{\rm C}}{10 R_\sun}\right)^{-1} \left(\frac{L_{\rm S}}{10^4 L_\sun}\right)^{-1} \text{yrs}
\end{align}
that the secondary spends on the Hertzsprung gap after exhausting the
hydrogen in its core (i.e., the Kelvin-Helmholtz timescale of the
core).  Since our fiducial scenarios require the secondary to fill its
Roche lobe prior to the second SN, one might expect tidal alignment to
always be efficient.  Substantial uncertainties remain in the model
however.  Stars with partially radiative envelopes may have longer
tidal-friction timescales
\cite{2006epbm.book.....E,2008ApJS..174..223B}, and the stellar core
may not efficiently couple to its envelope, as suggested by recent
Kepler observations of core-rotation rates \cite{2012Natur.481...55B}.
Therefore, for completeness, we also explore the ``extreme''
alternative scenario of completely inefficient tidal alignment.

Being dissipative in nature, tidal interactions decrease the
semimajor axis in addition to circularizing the orbit.  This change
is small compared to that induced by CE evolution, as discussed in the
next Section, and can therefore be neglected along with the orbital
changes produced by other phenomena (e.g. magnetic braking and mass
transfer).

\begin{figure}[htb]
\includegraphics[width=0.48\textwidth]{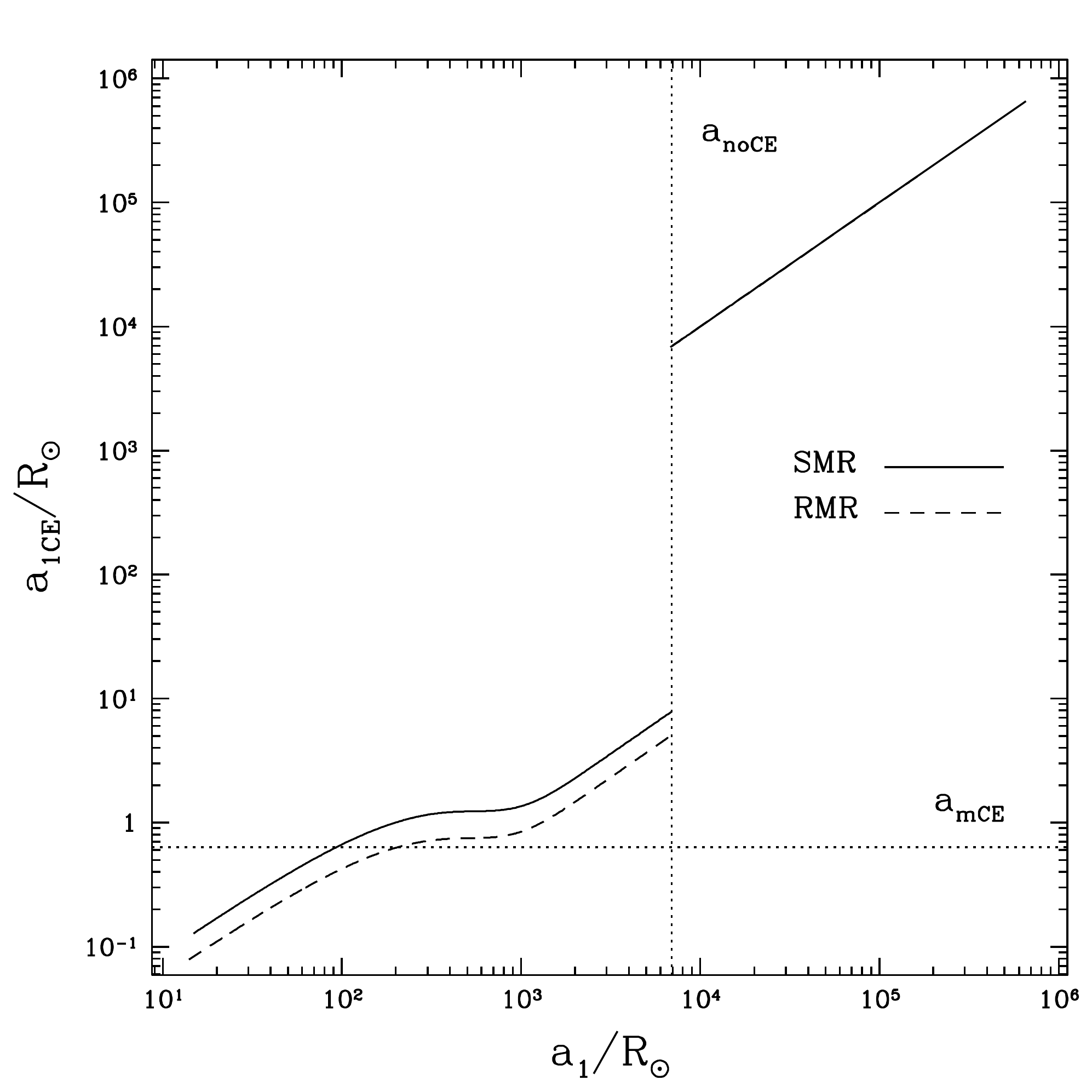}
\caption{The semimajor axis $a_{1CE}$ at the end of CE evolution as a
  function of its initial value $a_1$ in both the SMR and RMR
  scenarios.  If $a_1 > a_{\rm noCE}$, as given implicitly by
  Eq.~(\ref{E:noCE}), the secondary fails to fill its Roche lobe, no
  CE evolution occurs, and $a_{1CE} = a_1$.  If $a_{1CE} < a_{\rm
    mCE}$, as given implicitly by Eq.~(\ref{E:mCE}), the helium core
  of the secondary fills its Roche lobe prior to the end of CE and the
  binary merges, failing to eventually form a BH binary.  The nonlinear
  relationship between pre- and post-CE semimajor axes when
  $a_1<a_{\rm noCE}$ results from the nontrivial dependence of the CE
  efficiency parameter $\lambda$ on $a_1$, as given by
  Eq.~(\ref{lambdavar}).}
\label{CEev}
\end{figure}

\subsection{Common-envelope evolution} 
\label{A:CE}

If the semimajor axis $a_1$ of the binary following the first SN is
greater than $a_{\rm noCE}$, as determined from the constraint
\begin{equation} \label{E:noCE}
R_L(a_{\rm noCE}, M''_{Sf}, M'_{\rm BH}) = R^{\prime\prime}_G
\end{equation}
with $R^{\prime\prime}_G$ given by Eq.~(\ref{E:RG}), the secondary
does not fill its Roche lobe and no CE evolution occurs.  For smaller
values of $a_1$, we use conservation of energy to determine how much
the binary's orbit shrinks during CE evolution.  The gravitational
binding energy of the CE can be expressed as
\begin{align}
E_b=-\frac{GM''_{Sf}\left( M''_{Sf} - M''_C \right)}{\lambda \mathcal{R}}~,
\end{align}
where $M''_{Sf}$ is the mass of the secondary at the onset of CE
evolution, $M''_{Sf} - M''_C$ is the mass lost by the secondary during
this evolution, $\mathcal{R} = R_L(a_1, M''_{Sf}, M'_{\rm BH})$ is the
Roche-lobe radius of the secondary at the onset of CE evolution, and
$\lambda$ is a dimensionless parameter of order unity that depends on
the mass and structure of the secondary, notably the location of the
core-envelope boundary.  Full stellar-evolution codes can be used to
calculate the appropriate value of $\lambda$ for our BH progenitors
\cite{2010ApJ...716..114X,2010ApJ...722.1985X,2011ApJ...743...49L}. We
adopt an analytic fit to Fig.~3 of \cite{2012ApJ...759...52D}, which
summarizes the results of these calculations:
\begin{align}
\lambda= ae^{-b \,\mathcal R/ R_\sun}+c~,
\label{lambdavar}
\end{align}
where $a=0.358$, $b=7.19 \times 10^{-3}$, and $c=0.05$.  Conservation
of energy during CE evolution implies
\begin{equation}
-\frac{GM'_{\rm BH}M''_{Sf}}{2a_1} + E_b = -\frac{GM'_{\rm BH}M''_C}{2a_{1CE}}~;
\end{equation}
solving for $a_{1CE}$ yields
\begin{align}
a_{1CE} = a_1 \frac{M''_{\rm C}}{M''_{\rm Sf}}\left(1 + \frac{2}{ \lambda}\frac{a_1}{\mathcal R}
\frac{M''_{\rm Sf}-M''_{\rm C}}{M'_{\rm BH}} \right)^{-1}~.
\label{CEphase}
\end{align}
If $a_{1CE}$ is less than $a_{\rm mCE}$, as determined from the
constraint
\begin{equation} \label{E:mCE}
R_L(a_{\rm mCE}, M''_C, M'_{\rm BH}) = R^{\prime\prime}_C
\end{equation}
with $R^{\prime\prime}_C$ given by Eq.~(\ref{E:RC}), the helium core
of the secondary itself fills its Roche lobe before the end of CE
evolution.  This leads to a prompt merger, preventing the eventual
formation of a BH binary.  Our final prescription for $a_{1CE}$ as a
function of $a_1$ is shown in Fig.~\ref{CEev}.  CE evolution is
crucial to our model, shrinking the semimajor axis by a factor $\sim
10^3$ and thereby allowing the eventual BH binary to merge in less
than a Hubble time.

Motivated by hydrodynamical simulations
\cite{2008ApJ...672L..41R,2012ApJ...746...74R} and previous work on
binary evolution, we neglect accretion onto the primary BH during CE
evolution.  These studies suggest that the BH accretes at
substantially less than the Bondi-Hoyle rate during the evolution,
accumulating $\lesssim 0.1 M_\sun$ in mass.  Given this small change
in mass, we are justified in ignoring any resulting changes in the BH
spin \cite{1999MNRAS.305..654K}.  As noted in
Appendix~\ref{A:InitMass}, we also neglect the expansion of naked
helium stars, and therefore explicitly forbid a helium-star CE phase
\cite{2003ApJ...592..475I}.


%

\end{document}